\documentclass[epj,numbook,final]{svjour}

\usepackage{graphicx} 
\usepackage{amsmath,amssymb}
\usepackage{dcolumn}
\usepackage{subfig}
\usepackage{hyperref}

\newcommand{\eps}{\epsilon}

\newcommand{\Dk}{\Delta}
\newcommand{\cd}{{\cal D}x(\cdot)}
\newcommand{\dotx}{\dot{x}}
\newcommand{\doty}{\dot{y}}

\newcommand{\la}{\langle}
\newcommand{\ra}{\rangle}
\newcommand{\bx}{\bar{x}}
\newcommand{\bq}{\bar{q}}
\newcommand{\bS}{\bar{S}}
\newcommand{\bV}{\bar{V}}
\newcommand{\bdV}{\dot{\bar{V}}}
\newcommand{\N}{{\cal N}}
\newcommand{\ph}{\varphi}
\newcommand{\dph}{\dot{\varphi}}
\newcommand{\J}{{\cal J}}
\newcommand{\lam}{\lambda}
\newcommand{\al}{\alpha}
\renewcommand{\d}{\mathrm{d}}
\newcommand{\nn}{\nonumber}

\begin{document}

\title{Asymptotics of work distributions: The pre-exponential factor\thanks{Dedicated to Werner Ebeling on the occasion of his 75th birthday.}}

\author{D. Nickelsen \and A. Engel}

\institute{Universit\"{a}t Oldenburg, Institut f\"{u}r Physik, 26111
  Oldenburg, Germany}

\date{Received: date / Revised version: date} 


\abstract{
We determine the complete asymptotic behaviour of the work distribution in driven stochastic systems described by Langevin equations. Special emphasis is put on the calculation of the pre-exponential factor which makes the result free of adjustable parameters. The method is applied to various examples and excellent agreement with numerical simulations is demonstrated. For the special case of parabolic potentials with time-dependent frequencies, we derive a universal functional form for the asymptotic work distribution.}

\maketitle


\section{Introduction}\label{intro}
With the discovery of work \cite{Jar} and fluctuation \cite{ECM,GaCo} theorems in stochastic thermodynamics (for recent reviews see \cite{Seifert,EsvB}), the traditional emphasis of statistical mechanics on averages was extended to include also large deviation properties. Indeed, averages like the one appearing in the Jarzynski equality 
\begin{equation}\label{eq:JE}
 e^{-\beta F}=\langle e^{-\beta W} \rangle
\end{equation} 
are dominated by unlikely realization of the random variable (here the work $W$), and detailed information about the tail of the corresponding probability distribution is necessary to obtain an accurate result (see e.g. \cite{Collin}). By definition, rare realizations are difficult to get, and consequently, numerically and even more experimentally generated histograms seldom reach far enough into the asymptotic regime. It is therefore desirable to have some additional and independent information about the asymptotic behaviour of the relevant probability distributions. 

In the present paper we use the so-called method of optimal fluctuation to determine the asymptotics of work distributions in driven Langevin system. Special emphasis is put on the calculation of the pre-exponential factor that makes any fitting between histogram and asymptotics superfluous. By considering various examples, we show that the results for averages like (\ref{eq:JE}) improve significantly if the pre-exponential factor is included. We use a novel method \cite{KiMcKa} to determine this pre-factor which builds on the spectral $\zeta$-function of Sturm-Liouville operators. The\linebreak method is very efficient and straightforward in its numerical implementation. It also allows to handle the case of zero modes which is relevant in the present situation. For harmonic potentials with time-dependent frequency, we are able to derive the general form of the asymptotics of the work distribution analytically. 

The paper is organized as follows. Section \ref{sec:be} gives the basic equations and fixes the notation. In section \ref{sec:det} we recall the basic steps in the determination of functional determinants from spectral $\zeta$-functions and adapt the procedure to the present situation. Section \ref{sec:examples} discusses two examples, one that can be solved analytically and merely serves as a test of the method, and one for which the analysis has to be completed numerically. In section \ref{sec:bp} we elucidate the particularly interesting case of a harmonic oscillator with time-dependent frequency. Here, substantial analytic progress is possible. Finally, section \ref{sec:conc} contains some conclusions. Some more formal aspects of the analysis are relegated to the appendices A-C.


\section{Basic equations}\label{sec:be}
We consider a driven stochastic system in the time interval $0\leq t\leq T$ described by an overdamped Langevin equation of the form 
\begin{equation}
  \label{eq:LE}
  \dot{x}=-V'(x,t) +\sqrt{2/\beta}\; \xi(t)\; .
\end{equation}
The degrees of freedom are denoted by $x$, the time-depen\-dent potential $V$ gives rise to a deterministic drift, and $\xi(t)$ is a Gaussian white noise source obeying 
$\la\xi(t)\ra\equiv 0$ and $\la \xi(t)\xi(t')\ra=\delta(t-t')$. Derivatives with respect to $x$ are denoted by a prime, those with respect to $t$ by a dot. The system is coupled to a heat bath with inverse temperature $\beta$. The initial state $x(t=0)=:x_0$ of the system is sampled from the equilibrium distribution at $t=0$ 
\begin{equation}\label{eq:rho_0}
 \rho_0(x_0)=\frac{1}{Z_0}\exp(-\beta V_0(x_0))
\end{equation}
with $V_0(x):=V(x,t=0)$ the initial potential and 
\begin{equation}\label{eq:defZ_0}
 Z_0=\int \d x \exp(-\beta V_0(x))
\end{equation} 
the corresponding partition function. 

During the process, the system is externally driven and the potential changes from $V_0(x)$ to $V_T(x):=V(x,t=T)$ according to a given protocol. The work performed by the external driving depends on the particular trajectory $x(\cdot)$ the system follows and is given by \cite{Sekimoto} 
\begin{equation}
  \label{eq:defwork}
  W[x(\cdot)]=\int_0^{T} \!\!\d t\; \dot{V}(x(t),t)\; .
\end{equation}
Due to the random nature of $x(\cdot)$, also $W$ is a random quantity. According to the general rule of transformation of probability, its pdf is given by
 \begin{align}
  P(W)=&\int \frac{\d x_0}{Z_0}\; e^{-\beta V_0(x_0)}\!\!  \int \d x_T \nn\\ \label{eq:defPofW}
       &\!\!\times\int\limits_{x(0)=x_0}^{x(T)=x_T} \!\!  {\cal D}x(\cdot)\; p[x(\cdot)] \; 
       \delta(W- W[x(\cdot)])\;  .
\end{align}
The probability measure in trajectory space is \cite{ChDe} 
\begin{equation}\label{eq:defp}
 p[x(\cdot)]=\N \exp\Big(-\frac{\beta}{4}\int_0^{T} \!\! \d t \; \big(\dotx+V'(x,t)\big)^2\Big)\; ,
\end{equation}
where for mid-point discretization in the functional integral we have 
\begin{equation}\label{eq:defN}
 \N=\exp\Big(\frac{1}{2}\int_0^T \!\!\d t\; V''(x(t),t)\Big)\;  .
\end{equation}
Using the Fourier representation of the $\delta$-function in (\ref{eq:defPofW}), we find 
\begin{align}
  P(W)=&\,\N \int \frac{\d x_0}{Z_0}\int \d x_T\nn\\ \label{eq:PofW}
       &\!\!\times\int\frac{\d q}{4\pi/\beta} \!\!\!\int\limits_{x(0)=x_0}^{x(T)=x_T} \!\!\!\!{\cal
        D}x(\cdot)\;e^{-\beta S[x(\cdot),q]}
\end{align}
with the action 
\begin{equation}
  \label{eq:defS}
 S[x(\cdot),q]= V_0(x_0)+\!\int\limits_0^{T}\!\d t
   \Big[\frac{1}{4}(\dot{x}+V')^2+\frac{iq}{2}\dot{V}\Big]-\frac{iq}{2}W\, .
\end{equation}
The asymptotic behaviour of $P(W)$ may now be determined by utilizing the contraction principle of large deviation theory \cite{Touchette}. Roughly speaking, this principle stipulates that the probability of an unlikely event is given by the probability of its most probable cause \cite{Lifshitz,HaLa}. In the present context this means that whereas {\em typical} values of $W$ are brought about by a variety of different trajectories $x(\cdot)$, the {\em rare} values from the tails of $P(W)$ are predominantly realized by one particular path $\bx(\cdot)$ maximizing $P[x(\cdot)]:=\rho_0(x_0)p[x(\cdot)]$ under the constraint ${W=W[x(\cdot)]}$. A convenient way to implement this idea in the present context, is to evaluate the integrals in (\ref{eq:PofW}) by the saddle-point method. This is formally equivalent to considering the weak noise limit $\beta\to\infty$.

Let us therefore study expression (\ref{eq:PofW}) in the vicinity of a particular trajectory $\bx(\cdot)$ and a particular value $\bq$ of $q$. We put $x(t)=\bx(t)+y(t)$ and $q=\bq+r$ and expand up to second order in $y(\cdot)$ and $r$. After several partial integrations, we find
\begin{equation}
 S[x(\cdot),q]=\bS+ S_{\mathrm{lin}}+S_{\mathrm{quad}}+\dots
\end{equation}
 where 
\begin{align}
 \bS=&\,S[\bx(\cdot),\bq] \;, \\
 S_{\mathrm{lin}}=&\,\frac{1}{2}\biggl[(\bV'_0-\dot{\bx}_0)y_0 + (\bV'_T+\dot{\bx}_T)y_T\nn\\
     & - \int_0^T \!\! \d t\; (\ddot{\bx}+\bdV'-\bV'\bV'' - i\bq\,\bdV')\;y\nn\\
     & - ir\Big(W-\int_0^T\!\! \d t\; \bdV\Big)\biggr] \;, \\ \label{eq:Squad}
 S_{\mathrm{quad}}=&\,\frac{1}{4}\biggl[(\bV_0''y_0-\doty_0)y_0+(\bV_T''y_T+\doty_T)y_T\nn\\
     & + \int_0^T\!\! \d t\;  y\Big(-\!\frac{\d^2}{\d t^2}+\bV''^2+\bV'\bV'''-(1\!-\!i\bq)\bdV''\Big)y\nn\\
     & + 2ir\int_0^T\!\! \d t\; \bdV' y\biggr]\; .
\end{align}
Here, the notation $\bV:=V(\bx(t),t)$ and similarly for the derivatives of $V$ has been used. 

The most probable trajectory $\bx(\cdot)$ realizing a given value $W$ of the work is specified by the requirement that $S_{\mathrm{lin}}$ has to vanish for any choice of $y(\cdot)$ and $r$. We hence get the Euler-Lagrange equation (ELE) 
\begin{equation}\label{eq:ELE}
 \ddot{\bx}+(1-i\bq)\bdV'-\bV'\bV''=0
\end{equation}
together with the boundary conditions
\begin{equation}\label{eq:elebc}
 \dot{\bx}_0-\bV'_0=0, \qquad \dot{\bx}_T+\bV'_T=0\;  .
\end{equation}
From the term proportional to $r$ we find back the constraint 
\begin{equation}\label{eq:h12}
 W=\int_0^T\!\! \d t\; \bdV\; .
\end{equation}
Note that for given $W$ the solution for $\bx(\cdot)$ is usually unique and includes the optimal choice of  its initial and final point.

Neglecting contributions stemming from $S_{\mathrm{quad}}$, we arrive at the estimate 
\begin{equation}\label{eq:asy1}
 P(W)\sim\exp(-\beta \bS)
\end{equation}
giving the leading exponential term for the asymptotic behaviour of $P(W)$. It is solely determined by the optimal trajectory itself: Using the properties (\ref{eq:ELE}), (\ref{eq:elebc}) and (\ref{eq:h12}) of $\bx(\cdot)$ in (\ref{eq:defS}), one can show
\begin{align}
 \bS=&-\frac{W}{2}+\frac{1}{2}(\bV_T+\bV_0)-\frac{1}{4}(\bx_T\bV_T'+\bx_0\bV_0') \nn\\ \label{eq:resbS}
   &+\frac{1}{4}\int_0^T\!\!\!\d t\,\bV'(\bV'-\bx\bV'')
   +\frac{1-i\bq}{4}\int_0^T\!\!\!\d t\, \bx\,\bdV' \; .
\end{align} 

In many cases an improved estimate including the pre-exponential factor is desireable. To obtain it, also the neighbourhood of the optimal trajectory has to be taken into account. This is possible by retaining $S_{\mathrm{quad}}$ in the exponent and by performing the Gaussian integrals over $y_0, y_T, y(\cdot)$ and $r$.

In order to calculate 
\begin{equation}
 I:=\int\! \d y_0\!\int\! \d y_T\!\!\int\!\frac{\d r}{4\pi/\beta} \!\!\!\!
     \int\limits_{y(0)=y_0}^{y(T)=y_T} \!\!\!\!\!\!\!{\cal D}y(\cdot)\,
        e^{-\beta S_{\mathrm{quad}}[y(\cdot),r]} \;,
\end{equation}
we determine the eigenvalues $\lam_n$ and normalized eigenfunctions $\ph_n(t)$ of the operator $A$ defined by (cf.~(\ref{eq:Squad})) 
\begin{equation}\label{eq:defA}
 A:= -\frac{\d^2}{\d t^2}+(\bV'')^2+\bV'\bV'''-(1-i\bq)\bdV''
\end{equation}
together with the Robin-type boundary conditions
\begin{equation}\label{eq:bcA}
\bV_0''\ph_n(0)-\dph_n(0)=0, \qquad \bV_T''\ph_n(T)+\dph_n(T)=0\;.
\end{equation}
Next we expand 
\begin{equation}
 y(t)=\sum_n c_n \ph_n(t)
\end{equation}
and replace the integrations over $y_0, y_T$ and $y(\cdot)$ by integrations over the expansion parameters $c_n$ according to 
\begin{align}
 I=&\,\J\! \int\!\prod_n \frac{\d c_n}{\sqrt{4\pi/\beta}} \nn\\ \label{eq:efexp}
   &\!\!\times\!\!\int\!\!\frac{\d r}{4\pi/\beta}
    \exp\Big(\!-\frac{\beta}{4}\sum_n \lam_n c_n^2 - \frac{i\beta r}{2}\sum_n c_n d_n\Big)\, . 
\end{align}
Here, we have introduced the notation
\begin{equation}\label{eq:defd}
 d_n :=\int_0^T \d t\; \ph_n(t) \bdV'(t) \;,
\end{equation} 
and $\J$ is a factor stemming from the Jacobian of the transformation of integration variables. With this transformation being linear, the Jacobian is a constant; with  transformations between the eigenfunction systems of different operators being orthogonal, this constant cannot depend on the special form of $V$. In appendix \ref{sec:appB} we show by comparison with an exactly solvable case that 
\begin{equation}\label{eq:J}
 \J=\sqrt{\frac{8\pi}{\beta}}\;.
\end{equation}
Assuming that all eigenvalues are strictly positive, as necessary for $\bx(\cdot)$ being a minimum of $S[x(\cdot)]$, the $c_n$ integrals may be performed and we find 
\begin{equation}
 I=\sqrt{\frac{8\pi}{\beta}}\frac{1}{\sqrt{\prod_n \lam_n}}\int\!\frac{\d r}{4\pi/\beta}
  \exp\Big(-\frac{\beta r^2}{4}\sum_n \frac{d_n^{\,2}}{\lam_n}\Big)\; .
\end{equation}
Integrating finally over $r$, we are left with
\begin{equation}
  I=\frac{\sqrt{2}}{\sqrt{\prod_n \lam_n}}\;\frac{1}{\sqrt{\sum_n\frac{d_n^2}{\lam_n}}}\;  .
\end{equation}
Using 
\begin{equation}\label{eq:hbe}
 \prod_n \lam_n=\det A\quad \text{and}\quad \sum_n\frac{d_n^2}{\lam_n}=\la \bdV'|A^{-1}|\bdV'\ra \;,
\end{equation}
we finally get
\begin{equation}\label{eq:main}
 P(W)=\frac{\N\!\sqrt{2}}{Z_0}\,\frac{e^{-\beta\bS}}{\sqrt{\det A\,\la \bdV'|A^{-1}|\bdV'\ra}}
      \big(1+{\cal O}(1/\beta)\big)\, .
\end{equation}
This is the main result of this section. The same expression was obtained in \cite{Engel2009} from a discretization of the functional integral.


\section{Calculating determinants from spectral $\zeta$-functions}\label{sec:det}

The determination of the extremal action $\bS$ occurring in (\ref{eq:main}) requires the solution of the ELE (\ref{eq:ELE}). Although this can be done analytically only in a few exceptional cases, its numerical solution poses in general no difficulty. Similarly, the calculation of $\la \bdV'|A^{-1}|\bdV'\ra$ is rather straightforward. One solves the ordinary differential equation
\begin{equation}\label{eq:hode}
 A\, \psi(t)=\bdV'(t)
\end{equation}
with boundary conditions (\ref{eq:bcA}) and uses 
\begin{equation}\label{eq:resAm1}
 \la \bdV'|A^{-1}|\bdV'\ra=\int_0^T \d t\; \psi(t) \bdV'(t)\; .
\end{equation}

The determination of $\det A$ is somewhat more involved. We will use a method introduced recently \cite{KiMcKa}, see also \cite{KiLo}, building on the spectral $\zeta$-function of Sturm-Liouville operators. The essence of the method is contained in the relation 
\begin{equation}\label{eq:resdetA}
 \frac{\det A}{\det A_{\mathrm{ref}}}=\frac{F(0)}{F_{\mathrm{ref}}(0)}\;,
\end{equation}
where the eigenvalues $\lam_n$ of $A$ are given by the zeros of $F(\lam)$, and similarly $F_{\mathrm{ref}}(\lam)=0$ determines the eigenvalues of the reference operator $A_{\mathrm{ref}}$. Moreover, one has to ensure $F(\lam)/F_{\mathrm{ref}}(\lam)\to 1$ for $|\lam|\to \infty$. For operators of the type considered here, 
\begin{equation}
 A=-\frac{\d^2}{\d t^2}+g(t)\; ,
\end{equation}
with homogeneous Robin boundary conditions 
\begin{align}\label{eq:bc1}
 a\, \ph(t=0)+b\, \dph(t=0)&=0\;, \\\label{eq:bc2}
 c\, \ph(t=T)+d\, \dph(t=T)&=0\; ,
\end{align}
a convenient choice for $F(\lam)$ is \cite{KiMcKa} 
\begin{equation}\label{eq:defF}
 F(\lam)=c\chi_\lam(T)+d\dot{\chi}_\lam(T) \; ,
\end{equation}
where $\chi_\lam(t)$ is the solution of the initial value problem
\begin{equation}\label{eq:odedetA}
 A \chi_\lam(t)=\lam \chi_\lam(t),\qquad \chi_\lam(0)=-b,\; \dot{\chi}_\lam(0)=a\; .
\end{equation}
Note that $\chi_\lam(t)$ is defined for general $\lam$. The initial condition in (\ref{eq:odedetA}) is chosen such that $\chi_\lam(t)$ satisfies the boundary condition (\ref{eq:bc1}) at $t=0$ for all values of $\lam$. Only if $\lam$ coincides with one of the eigenvalues of $A$, $\lam=\lam_n$, the boundary condition at $t=T$ is satisfied as well. In this case $\chi_\lam$ is proportional to the eigenfunction $\ph_n$ corresponding to $\lam_n$. The roots of the equation $F(\lam)=0$ are therefore indeed the eigenvalues of $A$. 

In appendix \ref{sec:appA} we show that for the reference operator 
\begin{equation}
 A_{\mathrm{ref}}=-\frac{\d^2}{\d t^2}
\end{equation}
with boundary conditions (\ref{eq:bc1}), (\ref{eq:bc2}) one finds 
\begin{equation}\label{eq:resdetA0}
 \frac{\det A_{\mathrm{ref}}}{F_{\mathrm{ref}}(0)}=-\frac{2}{bd}\; .
\end{equation}
Hence, combining (\ref{eq:resdetA}) and (\ref{eq:resdetA0}), the determination of $\det A$ boils down to the solution of the initial value problem (\ref{eq:odedetA}). Numerically, this is again straightforward.


\section{Examples}\label{sec:examples}

\subsection{The sliding parabola}\label{sec:movpar}

The simplest example for the class of problems considered is provided by a Brownian particle dragged by a harmonic potential moving with constant speed \cite{MaJa,Wang+,ZoCo,Cohen} 
\begin{equation}\label{eq:defV1}
   V(x,t)=\frac{(x-t)^2}{2}\, .
\end{equation}
In this case, the full distribution $P(W)$ is known analytically \cite{MaJa,ZoCo}:
\begin{equation}\label{eq:Pexact}
  P(W)=\sqrt{\frac{\beta}{2\pi\sigma_W^2}}
     \exp\Big(-\beta\,\frac{(W-\sigma_W^2/2)^2}{2\sigma_W^2}\Big) 
\end{equation}
where 
\begin{equation}
  \sigma_W^2=2(T-1+e^{-T})\; .
\end{equation}
This example hence merely serves as a test of our method. To apply (\ref{eq:main}), we first note that 
\begin{equation}\label{eq:h1}
 \N=e^{T/2} \qquad\text{and}\qquad Z_0=\sqrt{\frac{2\pi}{\beta}}\; .
\end{equation}
Moreover, the ELE (\ref{eq:ELE}) is linear and can be solved analytically: 
\begin{equation}\nonumber
  \bx(t;W)=\frac1 2 (2t+e^{-t}-e^{t-T})
           -W\frac{(2-e^{-t}-e^{t-T})}{2(T+e^{-T}-1)}\, .
\end{equation}
Using this result in (\ref{eq:resbS}), we find
\begin{equation}\label{eq:h2}
  \bS=\frac{(W-(T+e^{-T}-1))^2}{4(T+e^{-T}-1)} \; .
\end{equation}
The operator $A$ defined in (\ref{eq:defA}) is, for the potential (\ref{eq:defV1}), given by
\begin{equation}
 A= -\frac{\d^2}{\d t^2}+1
\end{equation}
with boundary conditions
\begin{equation}
\ph_n(0)-\dph_n(0)=0, \qquad \ph_n(T)+\dph_n(T)=0\;.
\end{equation}
In order to determine $F(0)$, we have to solve the initial value problem
\begin{equation}
 -\ddot{\chi}_0+\chi_0(t)=0, \qquad \chi_0(0)=1,\; \dot{\chi}_0(0)=1\; .
\end{equation}
The solution is $\chi_0(t)=e^t$ implying (cf. (\ref{eq:defF})) $F(0)=2e^T$. Using (\ref{eq:resdetA}) and (\ref{eq:resdetA0}) we hence find 
\begin{equation}\label{eq:h3}
 \det A=4 e^T\;.
\end{equation}
Finally, $\bdV'=-1$, and in order to calculate $\la \bdV'|A^{-1}|\bdV'\ra$, we have to solve 
\begin{align}
 &\ddot{\psi}(t)-\psi(t)=1 \;, \nn\\
 &\psi(0)-\dot{\psi}(0)=0 , \qquad \psi(T)+\dot{\psi}(T)=0
\end{align}
which gives
\begin{equation}
 \psi(t)=\frac{1}{2}(e^{t-T}+e^{-t}-2)\;.
\end{equation}
Plugging this into (\ref{eq:resAm1}) yields
\begin{equation}\label{eq:h4}
 \la \bdV'|A^{-1}|\bdV'\ra=T+e^{-T}-1\; .
\end{equation}
Combining (\ref{eq:main}), (\ref{eq:h1}), (\ref{eq:h2}), (\ref{eq:h3}), and (\ref{eq:h4}), we find back (\ref{eq:Pexact}). Since $P(W)$ is Gaussian, the quadratic expansion around the saddle-point already reproduces the complete distribution, i.e. there are no higher order terms in (\ref{eq:main}). 


\subsection{The evolving double-well}\label{sec:sun}
As a more involved example, we discuss the time-depen\-dent potential proposed in \cite{Sun}
\begin{equation}
  \label{e:sun_V}
  V(x, t) = \al_1x^4 + \al_2(1-rt)x^2 \;.
\end{equation}
For $t<1/r$, the potential has a single minimum, for $t>1/r$, it evolves into a double-well. We consider the time interval $0<t<T = 2/r$ which places the transition at ${t=T/2}$. In contrast to the previous example, neither the work distribution, nor its asymptotics can be determined using solely analytical techniques. We will therefore generate the work distribution from simulations and solve the equations fixing the asymptotics numerically.

\begin{figure*}
  \subfloat[][]{\label{sf:sun_x}
	\includegraphics[width=0.9\columnwidth]{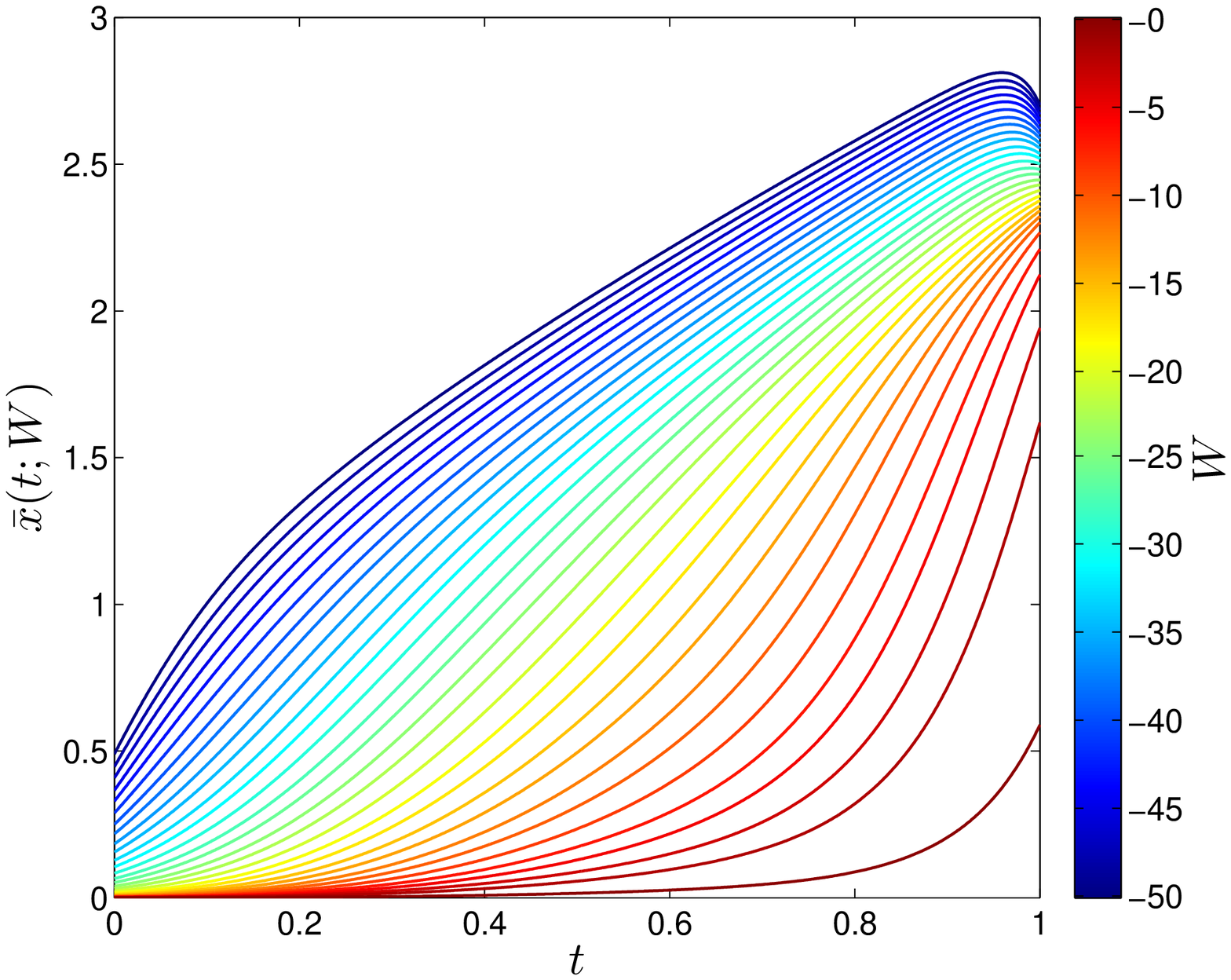}} \qquad
  \subfloat[][]{\label{sf:sun_iq}
	\includegraphics[width=0.9\columnwidth]{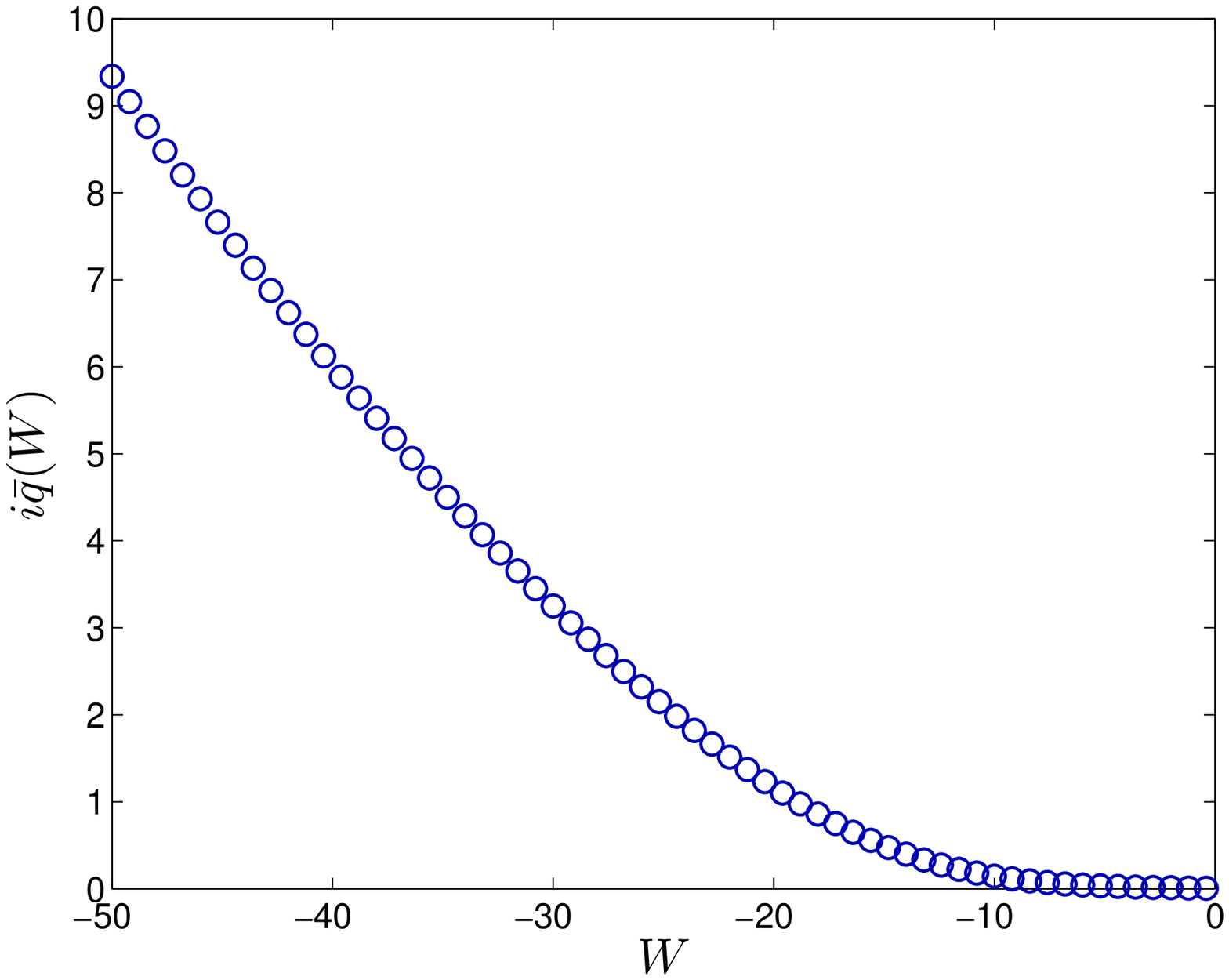}} \\ \\ \\
  \subfloat[][]{\label{sf:sun_detA}
	\includegraphics[width=0.9\columnwidth]{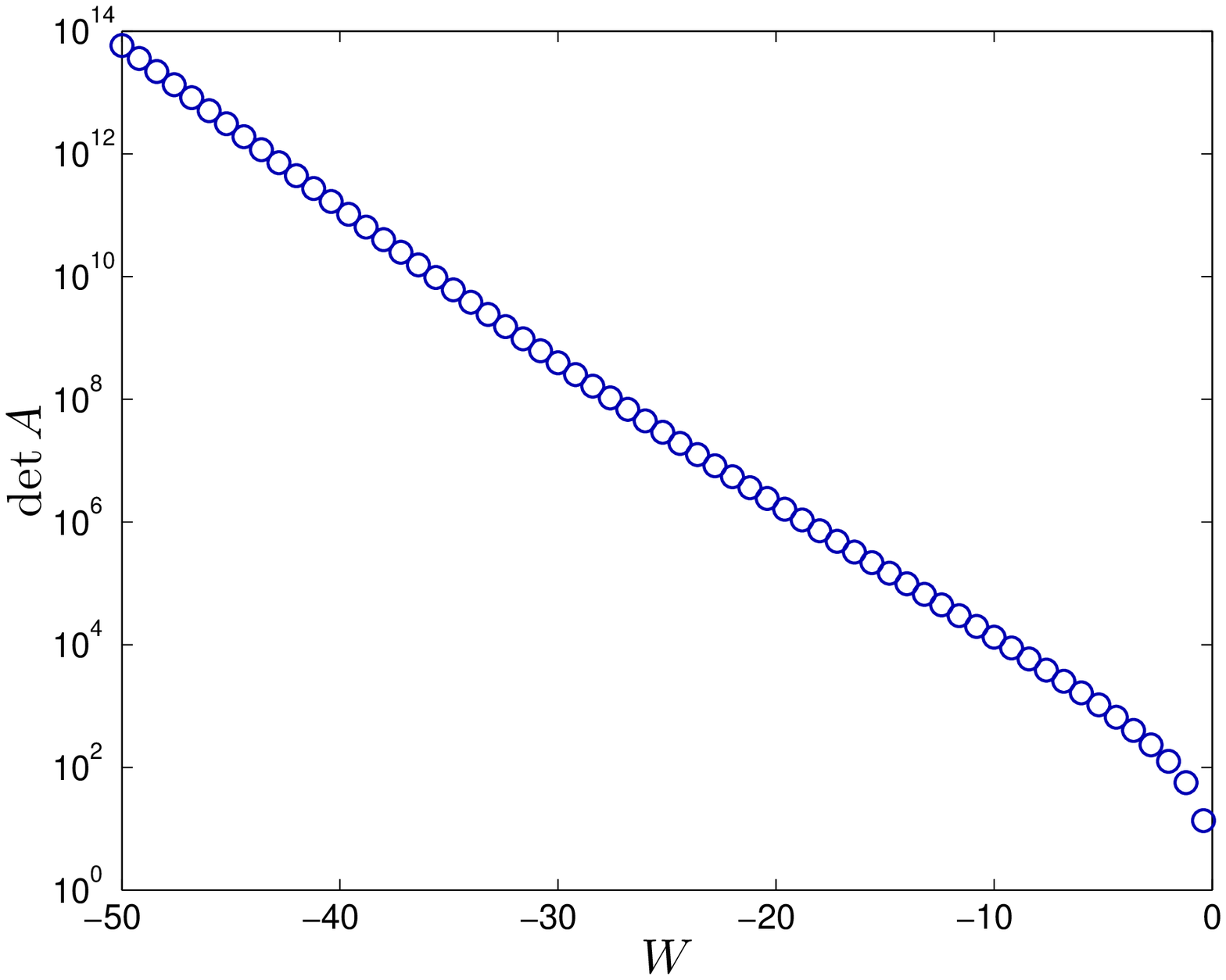}} \qquad
  \subfloat[][]{\label{sf:sun_R}
	\includegraphics[width=0.9\columnwidth]{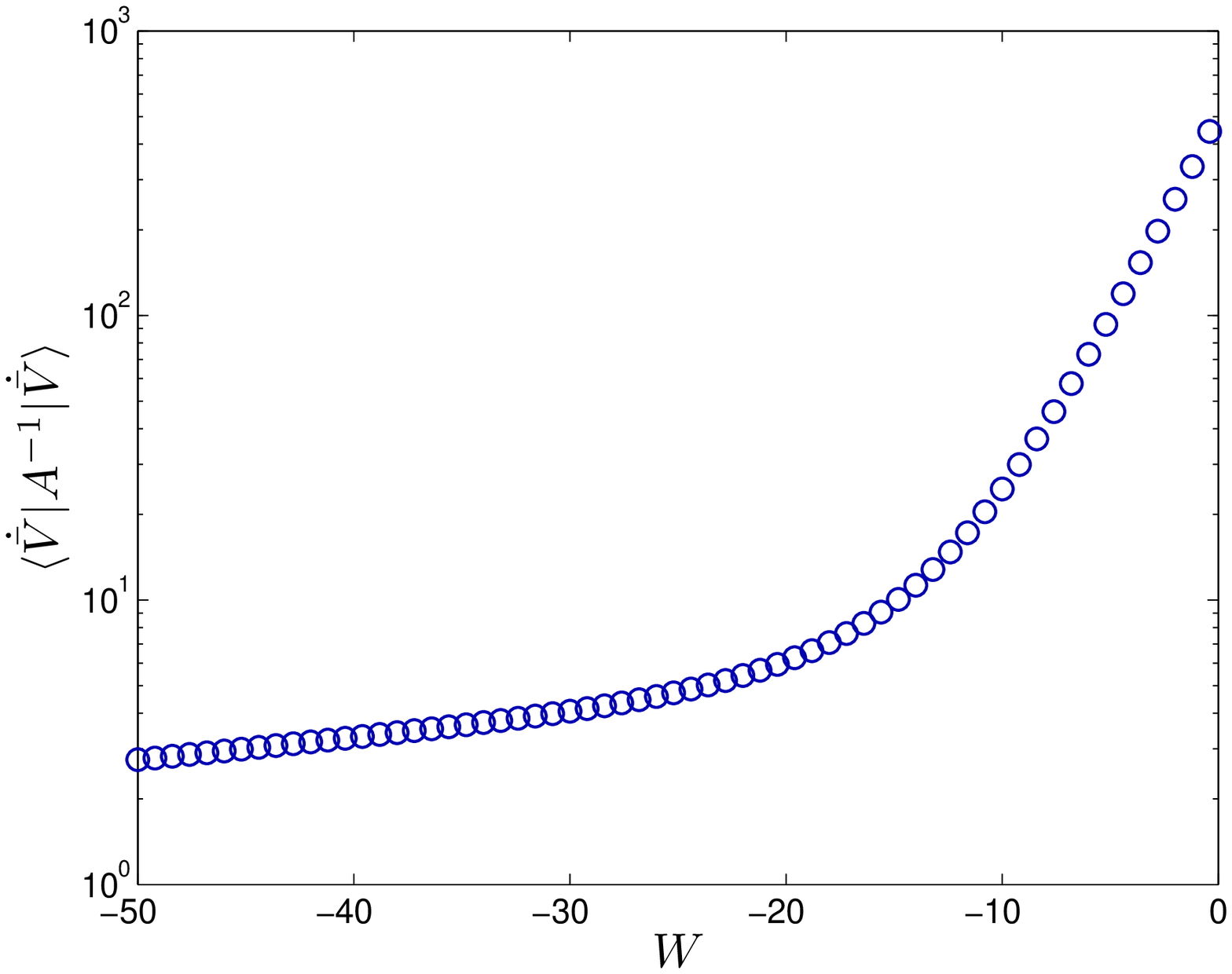}} \\ 
  \centering
  \caption{Numerical determination of the asymptotic work distribution for the evolving double well (\ref{e:sun_V}) with $\al_1 = 1/2$, $\al_2 = 6$, $T = 1$ and $\beta = 1$: (a) optimal trajectories $\bx(t;W)$, colours code values of the work $W$, (b) Lagrange parameter $i\bq$, (c)  determinant $\det A$ of the fluctuation operator $A$, and (d) quadratic form $\la\bdV'|A^{-1}|\bdV\ra$, all as function of $W$.}
  \label{f:sun_xiqdetAR}
\end{figure*}

\begin{figure*}
  \subfloat[][]{\label{sf:sun_PW_lin}
	\includegraphics[width=0.89\columnwidth]{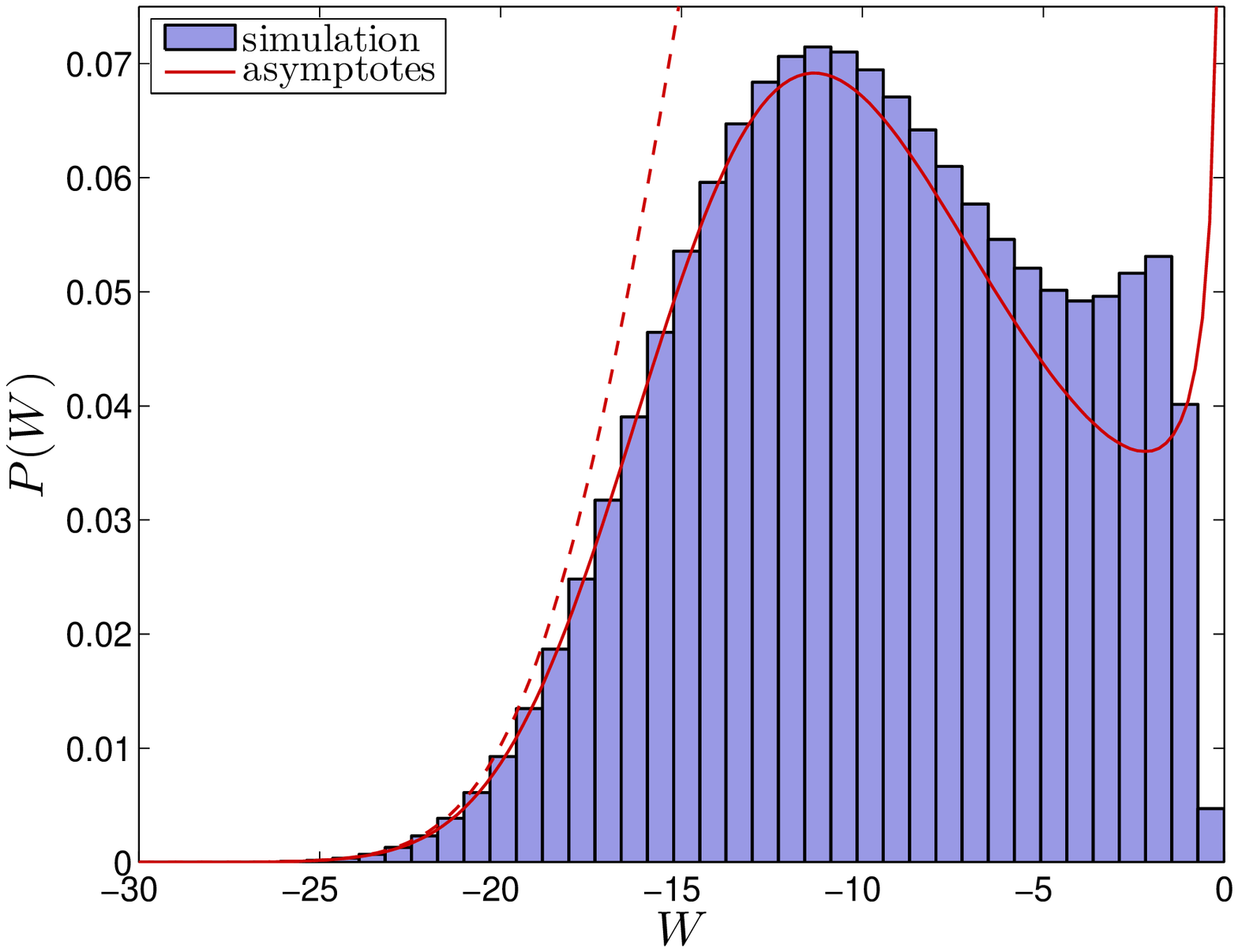}} \qquad
  \subfloat[][]{\label{sf:sun_PW_log}
	\includegraphics[width=0.9\columnwidth]{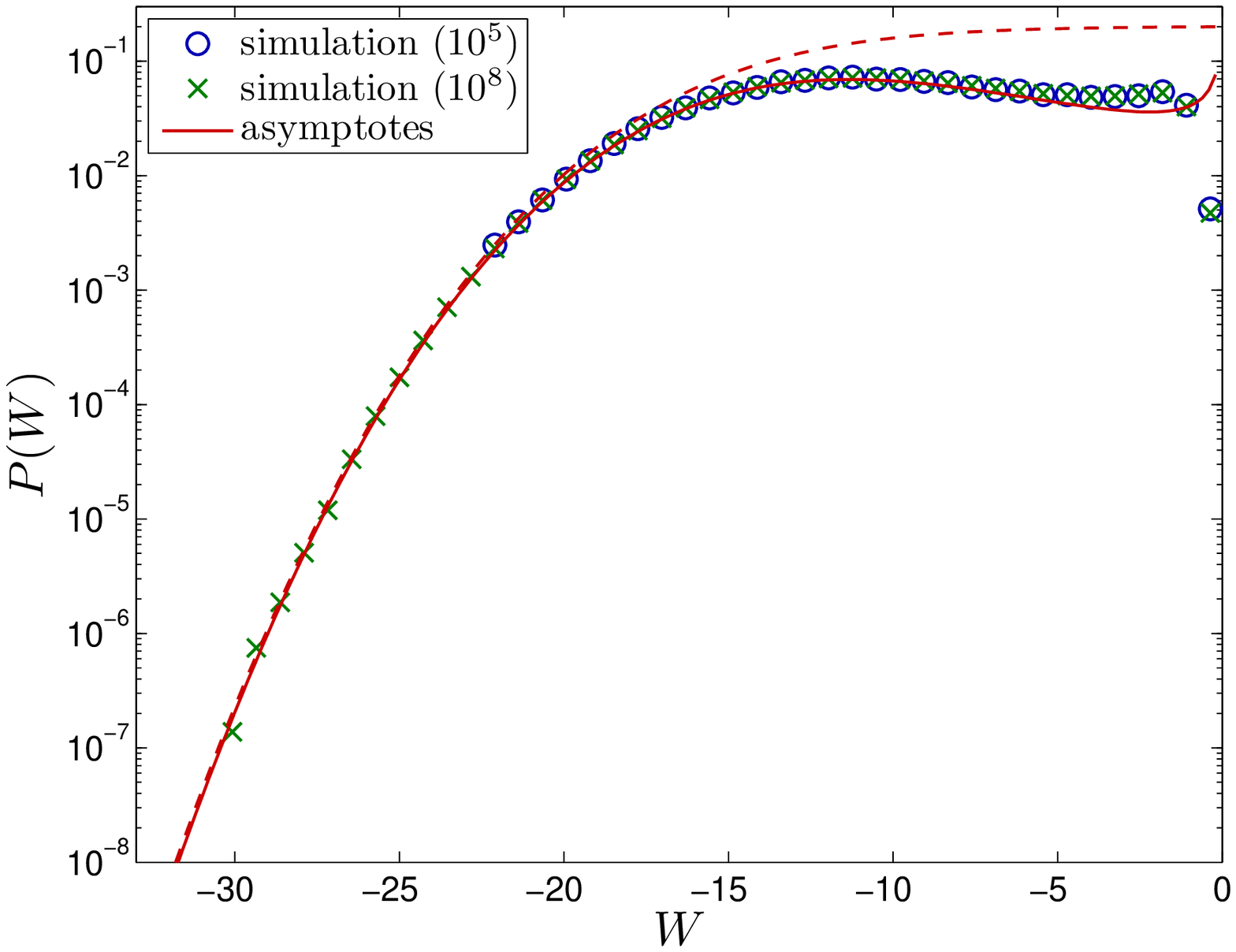}} \\ \\ \\
  \subfloat[][]{\label{sf:sun_PexpW_lin}
	\includegraphics[width=0.9\columnwidth]{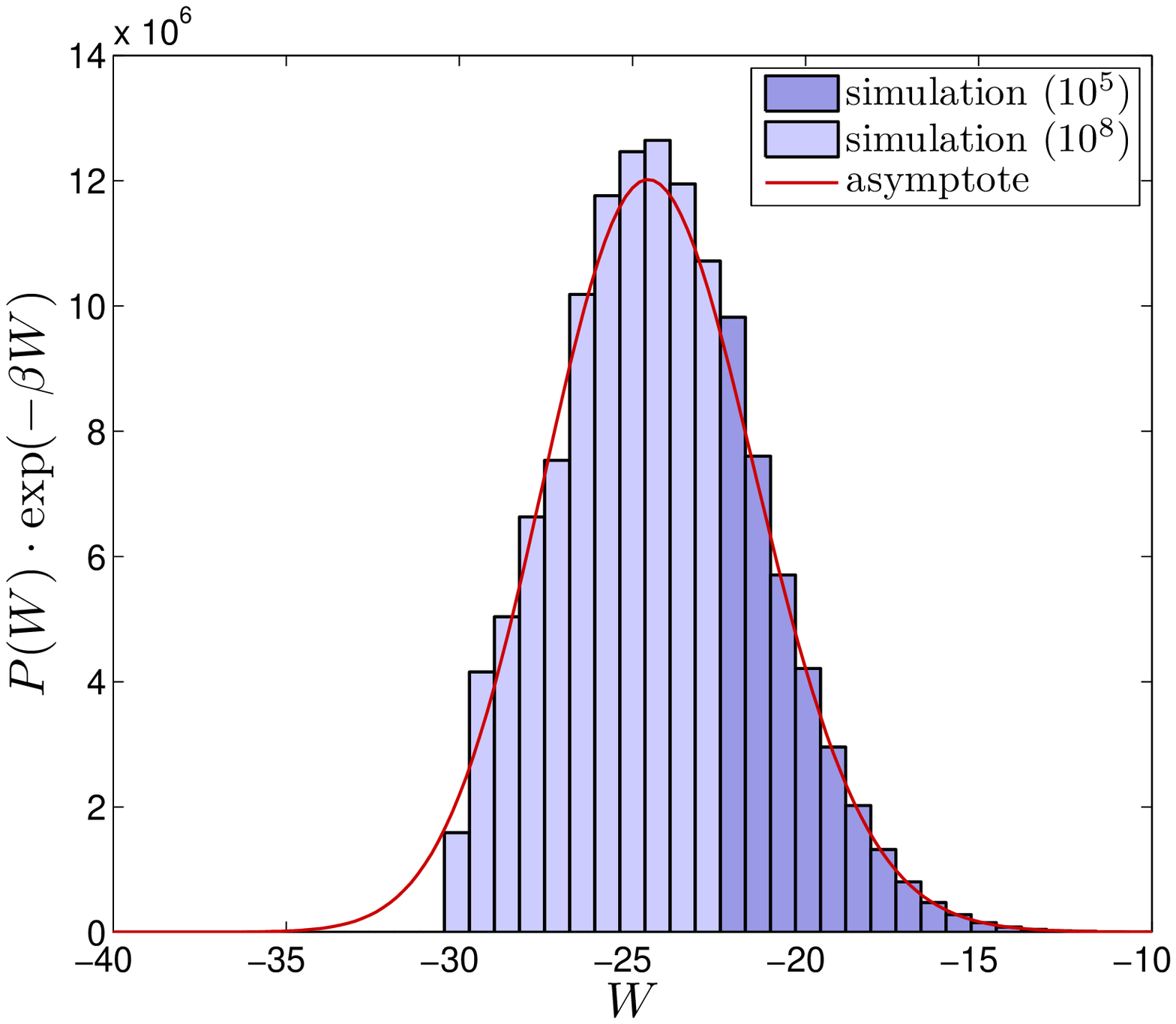}} \qquad
  \subfloat[][]{\label{sf:sun_PexpW_log}
	\includegraphics[width=0.89\columnwidth]{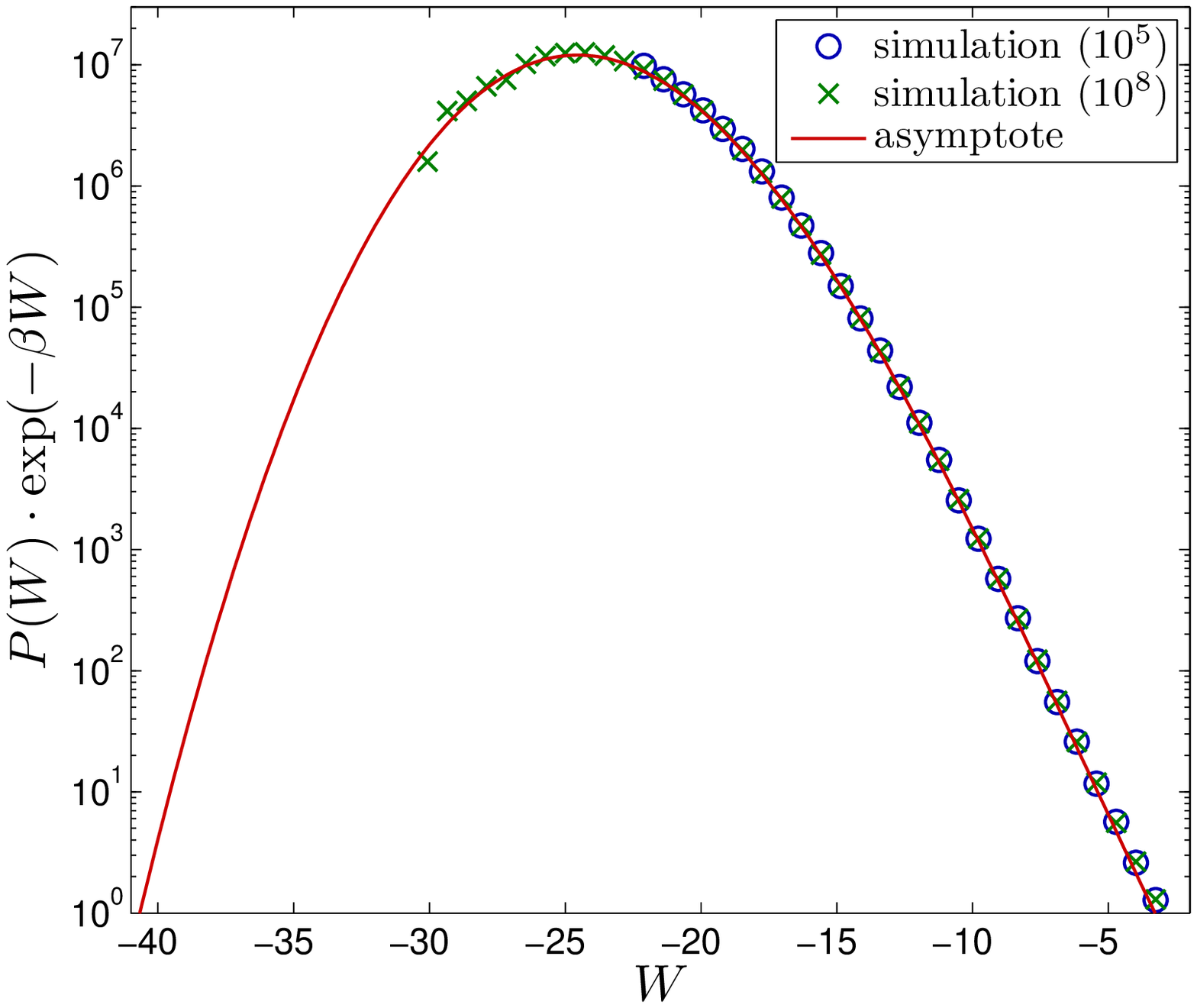}} \\
  \centering
  \caption{Work distribution for the evolving double-well (\ref{e:sun_V}) for $\alpha_1 = 1/2$, $\alpha_2 = 6$, $T = 1$ and $\beta = 1$. The histogram and the symbols show results from simulation of the Langevin dynamics (\ref{eq:LE}), the lines give the asymptotic forms (\ref{eq:resbp}) (full) with and (\ref{eq:asy1}) (dashed) without the pre-exponential factor. Subfigures (a) and (b) show a linear and logarithmic plot respectively of the work distribution itself, subfigures (c) and (d) display the distribution weighted with the factor $e^{-\beta W}$ as appearing, e.g., in the Jarzynski equality (\ref{eq:JE}). In (b) and (d) circles and crosses represent histograms based on $10^5$ and $10^8$ work values respectively, in (c) results from $10^8$ trajectories are shown in light blue.}
  \label{f:sun_PW}
\end{figure*}

To begin with, we have from (\ref{eq:defZ_0}) and (\ref{eq:defN})
\begin{align}
  Z_0 &= \int\!\!\d x\; \exp\big[-\beta(\al_1x^4+\al_2x^2)\big] \\
  \N &= \exp\big[(6\al_2x^2+\al_2)T-\frac{1}{2}\alpha_2rT^2\big] \;.
\end{align}
The ELE (\ref{eq:ELE}) reads 
\begin{align}
  \ddot\bx = &\, 48\al_1^2\bx^5 + 32\al_1\al_2(1-rt)\bx^3 \nn\\\label{e:sun_ele}  
  & + \big[4\al_2^2(1-rt)^2 + 2\al_2r(1-i\bq)\big]\bx \;,
\end{align}
and its boundary conditions (\ref{eq:elebc}) are of the form
\begin{align}  
  \dot\bx_0 &= 4\al_1\bx_0^3 + 2\al_2\bx_0 \;, \nn\\ \label{e:sun_elebc} 
  \dot\bx_T &= -4\al_1\bx_T^3 - 2\al_2(1-rT)\bx_T \;.
\end{align}
The constraint (\ref{eq:defwork}) is 
\begin{equation}
  \label{e:sun_constr}
  W = −\al_2 r \int_0^T\!\!\d t\;\bx(t;\bq)^2 \; .
\end{equation}
These equations can be solved numerically using a standard relaxation algorithm. The resulting optimal trajectories $\bx(t; W)$ and the corresponding Lagrange parameters $i\bq(W)$ are shown in Fig.~\ref{sf:sun_x} and Fig.~\ref{sf:sun_iq}, respectively. Due to the mirror symmetry of the potential, there are for each value of $W$ two optimal trajectories $\pm\bx(t; W)$, from which we only display the positive one. 
The operator $A$ from (\ref{eq:defA}) acquires the form
\begin{align}
  A = - \frac{\d^2}{\d t^2} &+ 240\al_1^2\bx^4 + 96\al_1\al_2(1-rt)\bx^2 \nn\\ \label{e:sun_A}
      &+ 2\al_2r(1-i\bq) + 4\al_2^2(1-rt)^2
\end{align}
with the boundary conditions (\ref{eq:bcA})
\begin{align}
  \big[12\al_1\bx_0^2+2\al_2\big]\ph_n(0)-\dot\ph_n(0) &= 0 \;, \nn\\
  \big[12\al_1\bx_T^2+2\al_2(1-rT)\big]\ph_n(T)+\dot\ph_n(T) &= 0 \;.
\end{align}
To obtain $\det A$ from (\ref{eq:resdetA}), we determine according to (\ref{eq:defF})
\begin{equation}
  \label{e:sun_F0}
  F(0) = \big[12\al_1\bx_T^2+2\al_2(1-rT)\big]\chi_0(T) + \dot\chi_0(T)
\end{equation}
by solving numerically the initial value problem (\ref{eq:odedetA})
\begin{align} 
 &\begin{aligned}
  \ddot\chi_0(t) = &\bigl[240\al_1^2\bx^4 + 96\al_1\al_2(1-rt)\bx^2 \nn \\
  & +2\al_2r(1-i\bq) + 4\al_2^2(1-rt)^2\bigr]\chi_0(t) = 0 \;,
 \end{aligned}
 \nn \\ \label{e:sun_chi0}
  &\dot\chi_0(0) = 1, \qquad \dot\chi_0(T) = 12\al_1\bx_T^2+2\al_2 \;. 
\end{align}
Note that this has to be done for each value of $W$ separately by using the appropriate results for  $\bx(t;W)$ and $\bq(W)$. The result for $\det A$ as a function of $W$ is depicted in Fig.~\ref{sf:sun_detA}. 

The last ingredient for the pre-exponential factor is $\la\bdV'|A^{-1}|\bdV\ra$ from (\ref{eq:resAm1}). To determine it, we need to solve the boundary value problem (\ref{eq:hode}), (\ref{eq:bcA})
\begin{align}
  &\begin{aligned}
     \ddot\psi(t)-\big[&240\al_1^2\bx^4 + 96\al_1\al_2(1-rt)\bx^2 \\
	 & + 2\al_2r(1-i\bq) + 4\al_2^2(1-rt)^2\big]\psi(t) + 2\al_2r\bx
   \end{aligned} \nn\\
  &\dot\psi(0)=\big(12\al_1\bx_0^2+2\al_2\big)\psi(0) \nn\\ \label{e:sun_psi}
  &\dot\psi(0)=-\big[12\al_1\bx_T^2+2\al_2(1-rT)\big]\psi(T) 
\end{align}
for each $\bx(t;W)$ and $\bq(W)$ and use the result in (\ref{eq:resAm1})
\begin{equation}
  \label{e:sun_R}
  \la\bdV'|A^{-1}|\bdV\ra = -2\al_2r \int_{0}^{T}\!\!\d t\; \bx(t) \, \psi(t) \;.
\end{equation}
The values for $\la\bdV'|A^{-1}|\bdV\ra$ obtained in this way are shown in Fig.~\ref{sf:sun_R}. 

Plugging the numerical results for $\N$, $Z_0$, $\bx$, $i\bq$, $\det A$ and $\la\bdV'|A^{-1}|\bdV\ra$ into (\ref{eq:main}) and adding an additional factor 2 to account for the two equivalent solutions $\pm\bx(t;W)$ for each value of $W$, we obtain the final result for the asymptotic form of the work distribution. 

To investigate the accuracy of this result, we employed the Heun scheme to simulate the Langevin equation (\ref{eq:LE}). In Fig.~\ref{sf:sun_PW_lin} we show the resulting histogram of $10^8$ work values and the asymptotic behaviour determined above. The dashed lines represent the incomplete asymptotic form (\ref{eq:asy1}) without pre-exponential factor, whereas the full line shows the complete asymptotics. In the former case an overall constant factor has to be adjusted, in the latter no free parameters remain. If the region of work values $-30<W<-20$ accessible from the simulation using $10^8$ trajectories is utilized for the fit in the incomplete asymptotics, the two asymptotic expressions almost coincide in the tail of the distribution. Away from the asymptotic regime, however, they differ markedly from each other. If, therefore, less data would be available, the fitted incomplete asymptotics could badly fail to reproduce the true asymptotic behaviour, see also Fig.~\ref{sf:atmpar_PW_log}. The parameter-free complete asymptotics is clearly advantageous.

Furthermore, there is a broad range of excellent agreement between histogram and complete asymptotics. This becomes in particular apparent when examining the \linebreak weighted work  distributions $P(W)\exp(-\beta W)$ shown in  Fig.~\ref{sf:sun_PexpW_lin} and  Fig.~\ref{sf:sun_PexpW_log}. The average $\langle\exp(-\beta W)\rangle$ appearing in the Jar\-zynski equality (\ref{eq:JE}) could already be accurately determined without the histogram at all by using nothing more than the complete asymptotics of $P(W)$.


\section{The breathing parabola} \label{sec:bp}

A particularly interesting class of examples is provided by harmonic oscillators with time dependent frequency \cite{Jar3,Carberry+}
\begin{equation}\label{eq:brepa}
  V(x,t)=\frac{k(t)}{2}\,x^2\; .
\end{equation}
Except for some special choices of $k(t)$, the full pdf of work is not known analytically. For our purpose the case of a monotonously decreasing function $k(t)$ is most appropriate. Then $W\leq 0$ and we aim at determining the asymptotic form of $P(W)$ for $W\to-\infty$. 

The ELE (\ref{eq:ELE}) is given by  
\begin{equation}
  \label{eq:ELE2}
  \ddot{\bx}+\big((1-i\bq)\dot{k}-k^2\big)\bx=0
\end{equation}
whereas the boundary conditions (\ref{eq:elebc}) acquire the form
\begin{equation}\label{eq:bcbp}
  \dot{x}_0=k_0\, x_0 \qquad\text{and}\qquad \dot{x}_T=-k_T\, x_T \;.
\end{equation}
These equations constitute themselves a Sturm-Liouville eigenvalue problem. Consequently, there are {\em infinitely}\linebreak {\em many} values $\bq^{(0)}, \bq^{(1)}, ...$ for $\bq$. Due to the mirror symmetry of the potential, each value $\bq^{(n)}$ again admits two non-trivial solutions $\pm \bx^{(n)}(\cdot)$. 

The somewhat unusual feature of this situation is that the different solutions $\bq^{(n)}$ are not related to the value $W$ of the constraint. Also, the functional form of $\bx^{(n)}(\cdot)$ is independent of $W$. The connection with the work value $W$ under consideration is brought about exclusively by the prefactor of $\bx^{(n)}(\cdot)$ which in view of (\ref{eq:defwork}) and (\ref{eq:brepa}) must be $\sqrt{|W|}$. 

For the saddle-point approximation in (\ref{eq:PofW}) this means that for any value of $W$ there are infinitely many stationary points of  $P[x(\cdot)]$. Moreover, from (\ref{eq:defA}) and (\ref{eq:bcA}) we find 
\begin{align}
 &A=-\frac{\d^2}{\d t^2} -(1-i\bq)\dot{k} +k^2\;, \\  
 &k_0\, x_0 - \dotx_0=0\;,\qquad k_T\, x_T + \dotx_T=0\; .
\end{align} 
Comparing this with (\ref{eq:ELE2}), (\ref{eq:bcbp}) it is seen that the fluctuations around the optimal path $\bx(\cdot)$ are governed by the same operator as the optimal path itself, as usual for a quadratic action. Together with the homogeneous boundary conditions (\ref{eq:elebc}), this implies that every saddle-point $\{\bq^{(n)},\pm\bx^{(n)}(\cdot)\}$ has a {\em zero mode}, namely the optimal path $\bx^{(n)}(\cdot)$ itself. From Courant's nodal theorem we know that $\bx^{(n)}(\cdot)$ has $n$ nodes. But if the Hessian of the saddle-point $\{\bq^{(n)}, \pm\bx^{(n)}(\cdot)\}$ has the eigenfunction $\bx^{(n)}(\cdot)$ with zero eigenvalue and $n$ nodes, it consequently must have $(n-1)$ eigenfunctions with {\em negative} eigenvalues. Therefore, only the solutions $\{\bq^{(0)}, \pm\bx^{(0)}(\cdot)\}$ of the ELE correspond to {\em maxima} of $P[x(\cdot)]$, all other solutions are saddle-points with unstable directions. These solutions are irrelevant for the asymptotics of $P(W)$, and it is therefore sufficient to determine the solutions to (\ref{eq:ELE2}), (\ref{eq:bcbp}) with the smallest value $\bq^{(0)}$of $\bq$. To lighten the notation, we will denote the corresponding solutions in the following simply by $\{\bq,\pm\bx(\cdot)\}$.

The general expression (\ref{eq:resbS}) for $\bS$ greatly simplifies for a parabolic potential. First $xV'=2V$, and the second and the third term in (\ref{eq:resbS}) cancel. Similarly, $xV''=V'$ and the forth term vanishes. Finally, $x\dot{V}'=2\dot{V}$ and the last term becomes proportional to $W$. We hence find the compact expression
\begin{equation}\label{eq:sbbp}
     \bS=\frac{i\bq}{2}\; |W|\; .
\end{equation}

A complication also arises in the determination of the pre-exponential factor. Here we cannot use (\ref{eq:main}) because the zero-mode makes $\det A=0$ and hence $A^{-1}$ becomes singular. Nevertheless, we may proceed as in section \ref{sec:be} up to eq.~(\ref{eq:efexp}) which now reads
\begin{align}
 I=&\,\sqrt{\frac{8\pi}{\beta}} \int\prod_n \frac{\d c_n}{\sqrt{4\pi/\beta}} \nn \\
  &\!\times\!\!\int\!\!\frac{\d r}{4\pi/\beta}\exp\Big(-\frac{\beta}{4}\sum_{n\geq 1} \lam_n c_n^2 - \frac{i\beta r}{2}\sum_n c_n d_n\Big)\,.
\end{align}
Integrating over $c_n$ with $n\geq 1$ yields 
\begin{align}
 I=&\sqrt{\frac{8\pi}{\beta}}\frac{1}{\sqrt{\det A'}} \int\!\frac{\d c_0}{\sqrt{4\pi/\beta}} \nn\\
   &\times\int\!\frac{\d r}{4\pi/\beta}\exp\Big(-\frac{\beta r^2}{4}\sum_{n\geq 1}\frac{d_n^{\,2}}{\lam_n}
    -\frac{i\beta r}{2}c_0 d_0\Big)\; ,
\end{align}
where we have used the usual notation 
\begin{equation}
 \det A':=\prod_{n\geq 1} \lam_n 
\end{equation} 
for a determinant omitting the zero mode. The remaining integrals over $r$ and $c_0$ are Gaussian and give
\begin{equation}
  I=\frac{\sqrt{2}}{\sqrt{d_0^2\; \det A'}}\; ,
\end{equation}
so that we end up with 
\begin{equation}\label{eq:mainmod}
 P(W)=2\;\frac{\N\sqrt{2}}{Z_0}\;\frac{e^{-\beta\bS}}{\sqrt{d_0^2\;\det A'}}\;
      \big(1+{\cal O}(1/\beta)\big)\; ,
\end{equation}
where the leading factor of 2 again accounts for the two equipollent saddle-points $\{\bq,\pm\bx(\cdot)\}$. Comparing this result with (\ref{eq:main}), we realize that in the presence of a zero-mode  we have to replace $\det A\;\la \bdV'|A^{-1}|\bdV'\ra$ by $\det A'\; d_0^2$. In view of (\ref{eq:hbe}), this is quite intuitive: With $\lam_0$ tending to zero, $\la\bdV'|A^{-1}|\bdV'\ra$ becomes more, and more dominated by $d_0^2/\lam_0$ and cancelling $\lam_0$ between $\det A$ and $\la \bdV'|A^{-1}|\bdV'\ra$ leaves us with (\ref{eq:mainmod}). 

We may finally express $d_0$ and $\det A'$ in terms of $\bx$. The former is calculated from its definition (\ref{eq:defd}) and $\ph_0=\bx/\|\bx\|$: 
\begin{equation}
 d_0=\int_0^T\!\!\! \d t\; \ph_0\bdV'=\int_0^T\!\!\! \d t\; \frac{\bx}{\|\bx\|}\dot{k}\bx=
       \frac{2W}{\|\bx\|}\; .
\end{equation} 
The determination of $\det A'$ may be accomplished by implementing a slight variation of the method described in section \ref{sec:det} \cite{KiMcKa}. As shown in appendix \ref{sec:appC}, eq.~(\ref{eq:resdetA}) has to be replaced by 
\begin{equation}\label{eq:resdetAmod}
 \frac{\det A'}{\det A_{\mathrm{ref}}}=\frac{\tilde{F}(0)}{F_{\mathrm{ref}}(0)}
\end{equation}
where 
\begin{equation}\label{eq:resFmod}
 \tilde{F}(0)=\frac{d \; \|\chi_0\|^2}{\chi_0(T)}\; .
\end{equation} 
Using $d=1$ and $\chi_0=\bx/\bx(0)$, we find from (\ref{eq:resdetA0}) 
\begin{equation}
 \det A'=2 \frac{\|\bx\|^2}{\bx_0\,\bx_T}\;.
\end{equation} 
Therefore, (\ref{eq:mainmod}) may be written as 
\begin{equation}\label{eq:mainbp}
 P(W)=\frac{\cal N}{Z_0}
  \frac{\sqrt{\bx_0\,\bx_T}}{|W|}\;e^{-\beta\tfrac{i\bq}{2}|W|}\,
  \big(1+{\cal O}(1/\beta)\big)\; ,
\end{equation}
where 
\begin{equation}
 {\cal N}=\exp\Big(\frac{1}{2}\int_0^T \!\!\d t\; k(t)\Big)
\end{equation} 
and 
\begin{equation}
 Z_0=\sqrt{\frac{2\pi}{\beta k_0}}
\end{equation} 
are easily calculated. 

We hence find for parabolas with time dependent frequency the universal asymptotic form 
\begin{equation}
 P(W)\sim C_1\sqrt{\frac{\beta}{|W|}}\;e^{-\beta\, C_2\, |W|}
\end{equation}  
with only the constants $C_1$ and $C_2$ depending on the special choice for $k(t)$. Note also that in this case the solution of the ELE is sufficient to get the full asymptotics including the prefactor.

As a simple example we first discuss the case
\begin{equation}
 k(t)=\left\{\begin{array}{lll} k_0 & \quad\text{for} & \quad0\leq t\leq \tau\\
                        k_T & \quad\text{for} & \quad\tau< t\leq T 
              \end{array}\right.\; .
\end{equation}
Here, $P(W)$ may again be calculated exactly. The particle gains energy only at $t=\tau$, where its position is still distributed according to $\rho_0(x)$. With $\Dk=k_0-k_T>0$, we hence find
\begin{align}
 P(W)&=\int \d x_\tau\, \rho_0(x_\tau)\;\delta(W+\frac{\Dk}{2}x_\tau^2) \nn\\\label{eq:resbp1}
     &=\sqrt{\frac{\beta k_0}{\pi\Dk|W|}}\; \exp\Big(-\beta\frac{k_0}{\Dk}|W|\Big)\; .
\end{align} 
The ELE (\ref{eq:ELE2}) is of the form 
\begin{equation}
 \ddot{\bx}-(1-i\bq)\Dk\delta(t-\tau) \bx -k^2\bx =0
\end{equation} 
and has the solution
\begin{equation}
 \bx(t)=\left\{\begin{array}{lll} \sqrt{\frac{2|W|}{\Dk}}\,e^{\,k_0(t-\tau)} & \text{for} & 0\leq t\leq \tau\\
               \sqrt{\frac{2|W|}{\Dk}}\, e^{-k_T(t-\tau)}  & \text{for} & \tau\leq t\leq T 
              \end{array}\right.
\end{equation} 
where
\begin{equation}
 (1-i\bq)=-\frac{k_0+k_T}{\Dk}\; .
\end{equation} 
Using this result in (\ref{eq:mainbp}) and (\ref{eq:sbbp}), we find back (\ref{eq:resbp1}).

\begin{figure*}
  \subfloat[][]{\label{sf:atmpar_PW_lin}
	\includegraphics[width=0.88\columnwidth]{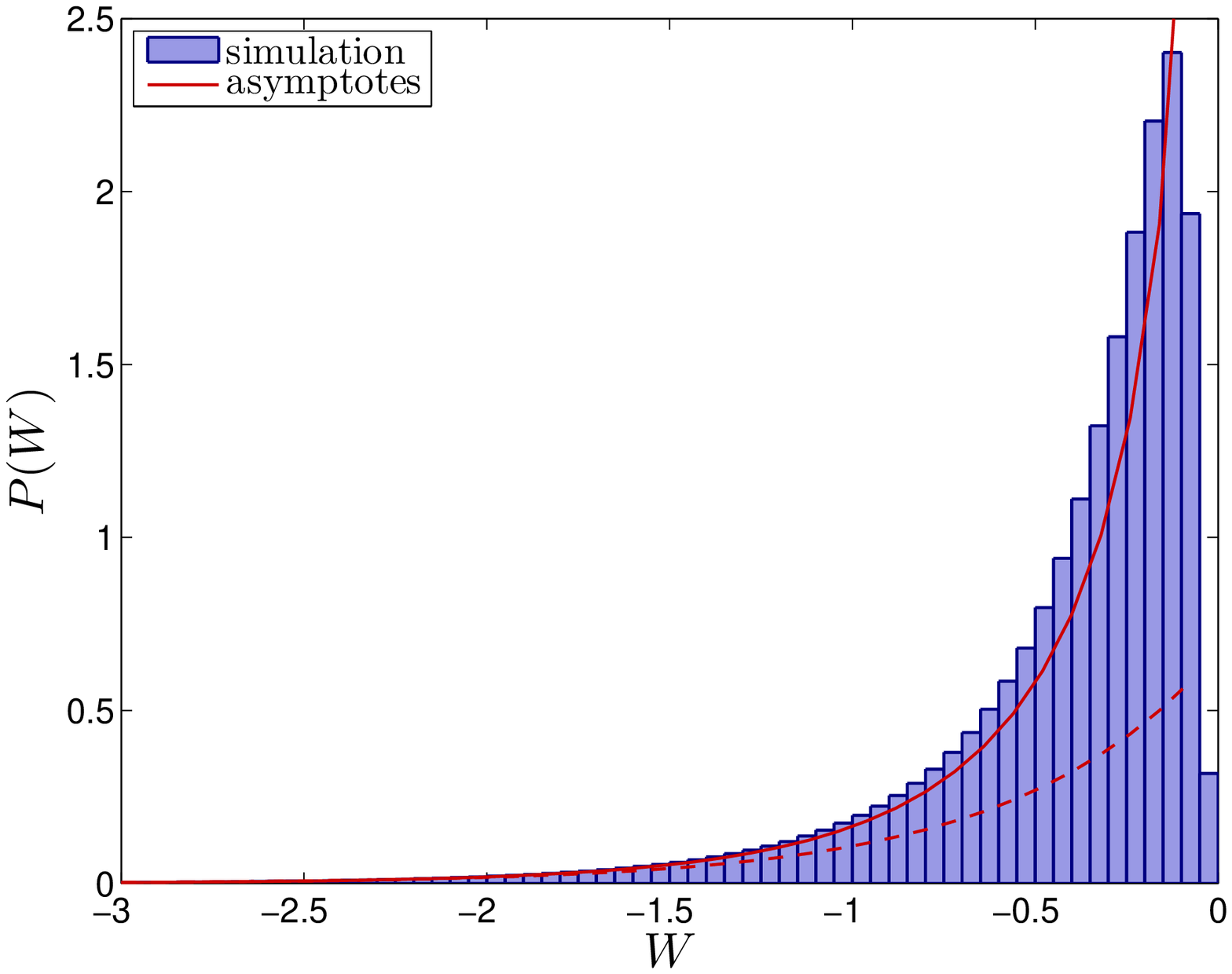}} \qquad
  \subfloat[][]{\label{sf:atmpar_PW_log}
	\includegraphics[width=0.9\columnwidth]{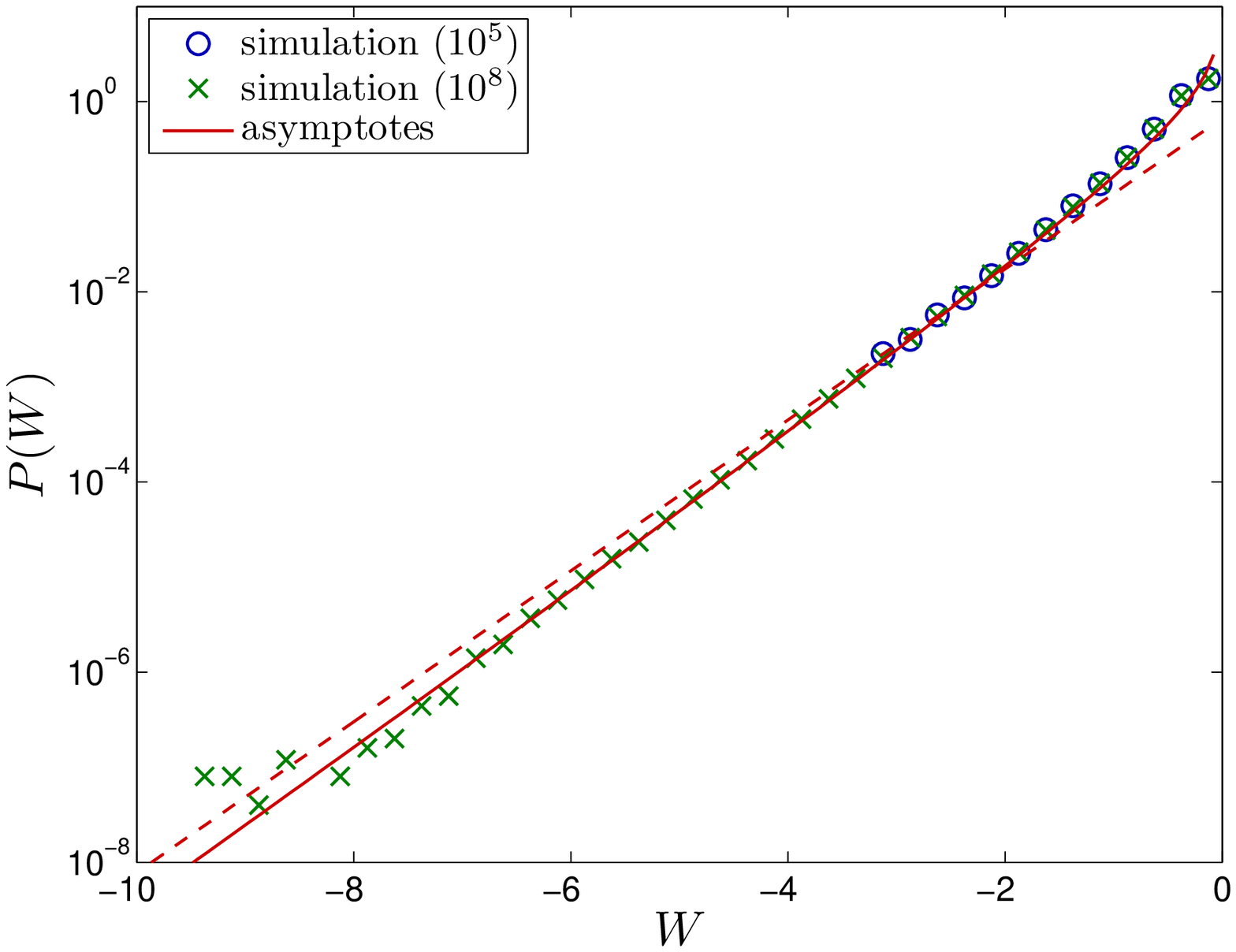}} \\ \\ \\
  \subfloat[][]{\label{sf:atmpar_PexpW_lin}
	\includegraphics[width=0.89\columnwidth]{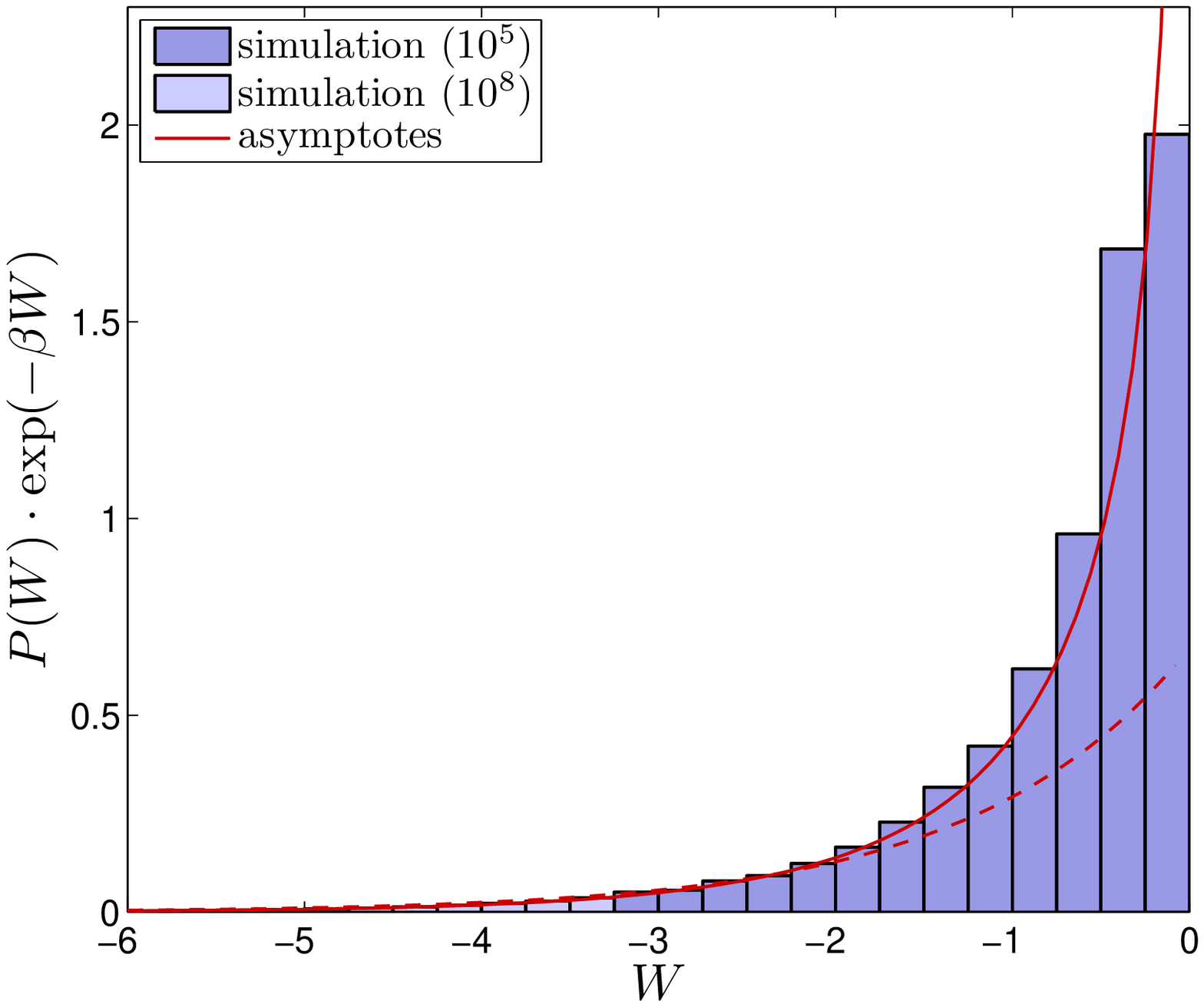}} \qquad
  \subfloat[][]{\label{sf:atmpar_PexpW_log}
	\includegraphics[width=0.9\columnwidth]{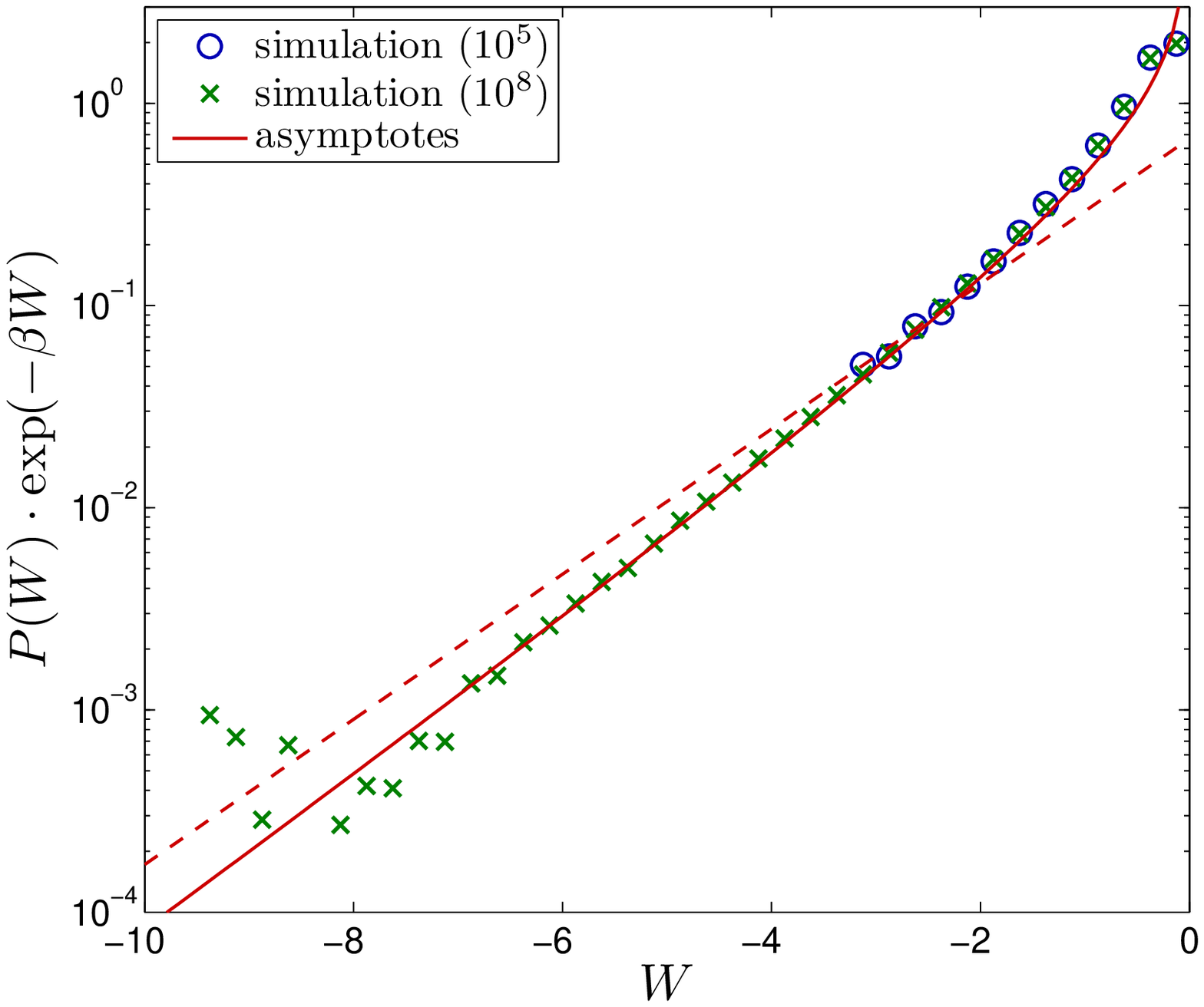}} \\
  \centering
  \caption{Work distribution for a breathing parabola (\ref{eq:brepa}) with protocol (\ref{eq:defk}) for $T = 2$ and $\beta = 1$. The histogram and the symbols show results from simulation of the Langevin dynamics (\ref{eq:LE}), the lines give the asymptotic forms (\ref{eq:resbp}) (full) with and (\ref{eq:asy1}) (dashed) without the pre-exponential factor. Subfigures (a) and (b) show a linear and logarithmic plot respectively of the work distribution itself, subfigures (c) and (d) display the distribution weighted with the factor $e^{-\beta W}$ as appearing, e.g., in the Jarzynski equality (\ref{eq:JE}). Circles and squares represent histograms based on $10^5$ and $10^8$ work values respectively, in (c) results from $10^8$ trajectories are shown in light blue.}
  \label{f:atmpar_PW}
\end{figure*}

A somewhat more general example is given by 
\begin{equation}  \label{eq:defk}
  k(t)=\frac{1}{1+t}\; .
\end{equation}
Here $P(W)$ is not known exactly, nevertheless, some analytical progress can be made in the determination of its asymptotics. To begin with, we have
\begin{equation}
 {\cal N}=\sqrt{1+T}\qquad\text{and}\qquad Z_0=\sqrt{\frac{2\pi}{\beta}}\; .
\end{equation} 
The ELE (\ref{eq:ELE2}) can be solved analytically with the result
\begin{align}
   \bx(t)=&\,\pm\frac{\sqrt{|W|}}{\sqrt{g(\mu)}}\sqrt{1+t} \nn\\ \label{eq:resELE}
          &\times\Big(2\mu\cos(\mu\ln(1+t))+\sin(\mu\ln(1+t))\Big)\; .
\end{align}
Here, with $\nu=2\mu\ln(1+T)$,
\begin{equation}\label{eq:resg}
   g(\mu)\!=\!\frac{1}{2}\Big[\big(\mu-\frac{1}{4\mu}\big)\sin\nu-\cos\nu
          +1+\nu\big(\mu+\frac{1}{4\mu}\big)\Big]>0\; ,
\end{equation}
and $\mu=\sqrt{i\bq-9/4}$ is the smallest root of
\begin{equation}
  \label{eq:EV}
  (4\mu2-3)\sin\frac{\nu}{2}-8\mu\cos\frac{\nu}{2}=0\; .
\end{equation}
This equation has to be solved numerically. The solution for $\mu$ yields the value of $i\bq$, determines the prefactor of $\bx$ in (\ref{eq:resELE}), and therefore fixes the complete asymptotics via (\ref{eq:mainbp}).

For the special case $T=2$ we find $\mu\cong 1.184$ implying $i\bq\cong 3.654$  which results into 
\begin{equation}\label{eq:resbp}
 P(W)\sim 1.021 \;\sqrt{\frac{\beta}{|W|}}\;\; e^{-1.827\, \beta\, |W|}\;.
\end{equation} 
In Fig.~\ref{f:atmpar_PW} we compare this asymptotics with results from numerical simulations of the Langevin equation. Here the incomplete asymptotics represented by the dashed line was fitted to the numerical results from $10^5$ realizations. It is clearly seen in the logarithmic plots (b) and (d) that this procedure does not yield reliable results for the far tail of the distributions. The full asymptotics including the pre-exponential factor is again clearly superior and describes the true distribution up to rather large values of $W$.


\section{Conclusion} \label{sec:conc}
In the present paper, we have shown how the complete asymptotic behaviour of work distributions in driven\linebreak Langevin systems may be determined. The calculation of the pre-exponential factor was accomplished by a method building on the spectral $\zeta$-function of the operator describing the quadratic fluctuations around the optimal trajectory. We have shown that the inclusion of the pre-exponential factor improves the asymptotics significantly and simplifies its application due to the absence of free parameters. For the examples considered, our results for the asymptotics match the outcome of extensive simulations perfectly and reach far into the region of work values accessible to simulations or experiments with moderate sample sizes. 

For the class of harmonic systems with time-dependent frequency, we established the universal form 
\begin{equation}\label{eq:conch}
 P(W)\sim C_1\sqrt{\frac{\beta}{|W|}}\;e^{-\beta\, C_2\, |W|}
\end{equation} 
of the asymptotic work distribution with only the two constants $C_1$ and $C_2$ depending on the detailed time-depen\-dence of the frequency. We note that this form is at variance with the findings of Speck and Seifert \cite{SpSe} who claim that the work distribution for Langevin systems with slow but finite driving must always be Gaussian. Our general result (\ref{eq:conch}) shows that while being presumably correct for the central part of the distribution, their statement does not hold for the asymptotics. We also note that the same asymptotic behaviour was found recently for a two-dimensional Langevin system with linear non-potential forces \cite{KNP}.

In view of the complex general expression (\ref{eq:resbS}) for the exponent in the asymptotics, in which all terms depend in a rather implicit way on the work value $W$, a further identification of universality classes for asymptotic work distributions remains an open challenge.


\begin{acknowledgement}
  We would like to thank Daniel Grieser for clarifying remarks on the determinant of Sturm-Liouville operators and Markus Niemann for helpful discussions.
\end{acknowledgement}


\appendix

\section{Calculation of ${\cal J}$}\label{sec:appB}
Consider as special case of (\ref{eq:LE}) the Langevin equation 
\begin{equation}
   \dot{x}=-c x +\sqrt{2/\beta}\; \xi(t)
\end{equation} 
with some real constant $c$. According to (\ref{eq:defp}) and (\ref{eq:defN}) the propagator of the corresponding Fokker-Planck equation is given by 
\begin{align}
 p(x_T,&T|x_0,0)= \nn\\
       &e^{cT/2} \!\!\!\!\int\limits_{x(0)=x_0}^{x(T)=x_T}\!\!\!\!\!\!\cd
        \exp\Big(-\frac{\beta}{4}\int_0^{T} \!\! \d t \; \big(\dotx+cx\big)^2\Big)\; .
\end{align} 
From the normalization condition ${\int \d x_T\; p(x_T,T|x_0,0)=1}$, we have 
\begin{align}
 1=&\,e^{cT/2} \sqrt{\frac{\beta(a+c)}{4\pi}}\int \d x_0 \exp\Big(-\frac{\beta(a+c)}{4}\,x_0^2\Big) \nn\\ \nn
 &\times\int \d x_T \!\!\!\! \int\limits_{x(0)=x_0}^{x(T)=x_T} \!\!{\cal D} x(\cdot) 
  \exp\Big(-\frac{\beta}{4}\int_0^{T} \!\! \d t \; \big(\dotx+cx\big)^2\Big)\; ,
\end{align} 
where $a>-c$ denotes some other real constant. By partial integration we find
\begin{align}
  1=&\,e^{cT/2} \sqrt{\frac{\beta(a+c)}{4\pi}}\int \d x_0 \int \d x_T \nn \\
    &\begin{aligned}
       \times\!\!\!\!\!\int\limits_{x(0)=x_0}^{x(T)=x_T} \!\!\!\!\!\!{\cal D} x(\cdot)
       \exp\Bigl(-\frac{\beta}{4}\biggl[(&ax_0-\dotx_0)x_0+(cx_T+\dotx_T)x_T \nn\\
       & + \int_0^{T} \!\! \d t \; x\big(-\frac{\d^2}{\d t^2} +c^2\big)x\biggr]\Bigr)
     \end{aligned} \nn\\ \label{eq:resJ}
   =&\,e^{cT/2}\; \sqrt{\frac{\beta(a+c)}{4\pi}}\; \J \;\frac{1}{\sqrt{\det A}}\;,
\end{align} 
where the operator $A$ is defined by 
\begin{align}
 &A \ph=-\ddot{\ph}+c^2\ph \;, \nn\\
 &a\,\ph(0)-\dot{\ph}(0)=0,\quad c\,\ph(T)+\dot{\ph}(T)=0\; .
\end{align} 
With the methods described in section \ref{sec:det} we easily find
\begin{equation}
 \det A=2(a+c)\;  e^{cT}\; ,
\end{equation} 
and comparison with (\ref{eq:resJ}) yields 
\begin{equation}
  \J=\sqrt{\frac{8\pi}{\beta}}\; .
\end{equation} 
Note that this value does not depend on $a$ and $c$ and is therefore valid for all situations considered in  the present paper. 


\section{Calculation of $\det A_{\mathrm{ref}}$ and $F_{\mathrm{ref}}(0)$}\label{sec:appA}
In this appendix we calculate the determinant of the operator  
\begin{equation}
 A_{\mathrm{ref}}=-\frac{\d^2}{\d t^2}
\end{equation}
with boundary conditions 
\begin{align}
 a\, \ph(t=0)+b\, \dph(t=0)&=0\\
 c\, \ph(t=T)+d\, \dph(t=T)&=0\; .
\end{align}
The main point for introducing this reference operator is that the ratio of determinants of two  Sturm-Liouville operators has much nicer analytic properties than each determinant individually \cite{KiLo}. Intuitively, this is related to the fact that in the naive interpretation of the determinant as product of all eigenvalues each determinant separately is clearly infinite, whereas their ratio may remain bounded. Our derivation closely follows \cite{KiLo}, where the corresponding analysis for Dirichlet boundary conditions was given.

We start with 
\begin{equation}\label{eq:ah0}
 \det A_{\mathrm{ref}}=e^{-\zeta_{\mathrm{ref}}'(0)}\; ,
\end{equation}
where 
\begin{equation}
 \zeta_{\mathrm{ref}}(s):=\sum_{n=1}^\infty \lam_{\mathrm{ref},n}^{\;-s}
\end{equation}
is the spectral $\zeta$-function of the operator $A_{\mathrm{ref}}$ with $\lam_{\mathrm{ref},n}$ denoting its eigenvalues \cite{KiMcKa}. Let $\chi_\lam(t)$ be the solution of 
\begin{equation}\label{eq:ah1}
 -\ddot{\chi}_\lam(t)=\lam \chi_\lam(t),\qquad \chi_\lam(0)=-b,\; \dot{\chi}_\lam(0)=a\;.
\end{equation}
Then
\begin{equation}
 F_{\mathrm{ref}}(\lam):= c \chi_\lam(T)+d \dot{\chi}_\lam(T)
\end{equation}
has zeros at all eigenvalues $\lam=\lam_{\mathrm{ref},n}$ of $A_{\mathrm{ref}}$. Correspondingly, $\d \ln F_{\mathrm{ref}}(\lam)/\d \lam$ has poles at these eigenvalues with residua given by their multiplicities. We may hence write 
\begin{equation}
 \zeta_{\mathrm{ref}}(s)=\frac{1}{2\pi i}\int_{\mathfrak{c}} \d\lam \, \lam^{-s}\frac{\d}{\d\lam}\ln F_{\mathrm{ref}}(\lam)\; ,
\end{equation} 
where the contour $\mathfrak{c}$ starts at $\lam=\infty+i\eps$, goes down parallel to the real axis, makes a half-circle around the lowest eigenvalue $\lam_{\mathrm{ref},1}>0$ and continues parallel to the real axis to $\lam=\infty-i\eps$. 

For the simple case of $A_{\mathrm{ref}}$ we may solve (\ref{eq:ah1}) and determine $F_{\mathrm{ref}}(\lam)$ explicitly:
\begin{equation}\label{eq:ha1}
 F_{\mathrm{ref}}(\lam)=(ad-bc)\cos\sqrt{\lam}T+(ca+bd\lam)\frac{\sin\sqrt{\lam}T}{\sqrt{\lam}}\; .
\end{equation} 
From this result we find 
\begin{equation}
 \frac{\d}{\d\lam}\ln F_{\mathrm{ref}}(\lam)\sim\sqrt{\lam}
\end{equation} 
for large $|\lam|$. Provided $s>1/2$, we may hence deform $\mathfrak{c}$ to the contour starting at $\lam=-\infty+i\eps$, going up to a half-circle around $\lam=0$ and running back to $-\infty-i\eps$. Taking into account that the integrand has a branch cut along the negative real axis and substituting $\lam=-x\pm i\eps$, we find (for $s>1/2$) 
\begin{align}
 \zeta_{\mathrm{ref}}(s)
  &=\frac{\sin\pi s}{\pi}\int_0^\infty \d x\, x^{-s}\frac{\d}{\d x}\ln F_{\mathrm{ref}}(-x)\\ \label{eq:ah2}
  &\begin{aligned}
     =\frac{\sin\pi s}{\pi}\!\int_0^\infty&\d x\, x^{-s} \\
     &\begin{aligned}
       \!\frac{\d}{\d x}\ln\!\Big[&(ad\!-\!bc)\cosh(\sqrt{x}t) \\
       & +\!(ca\!-\!bdx)\frac{\sinh\sqrt{x}t}{\sqrt{x}}\Big].
     \end{aligned}
   \end{aligned}
\end{align}
In order to use this expression in (\ref{eq:ah0}) it has to be analytically continued to $s=0$. No problems arise at the lower limit of integration, $x=0$. Using the representation
\begin{align}
 \frac{\d}{\d x}\ln F_{\mathrm{ref}}(-x)=&\,\frac{T}{2\sqrt{x}}+\frac{1}{2x} \\ \nn
  &\begin{aligned}
     +\frac{\d}{\d x}\ln\Big[&\frac{ad-bc}{2\sqrt{x}}(1+e^{-2\sqrt{x}T}) \\
     &+\frac{ca/x-bd}{2}(1-e^{-2\sqrt{x}T})\Big]\;,
   \end{aligned}
\end{align} 
we see, however, that the first two terms entail divergences at large $x$ for $s\to 0$. Splitting the integral in (\ref{eq:ah2}) at ${x=1}$, these dangerous terms may be integrated explicitly and the continuation to $s=0$ presents no further problems.

At the end we find
\begin{equation}
 \zeta_{\mathrm{ref}}'(s=0)=-\ln\left(\frac{2}{bd}(cb-ad-caT)\right)
\end{equation} 
implying
\begin{equation}\label{eq:detA0}
 \det A_{\mathrm{ref}}=\frac{2}{bd}(cb-ad-caT)\; .
\end{equation} 
To use this result in (\ref{eq:resdetA}), we have finally to determine $F_{\mathrm{ref}}(0)$. To this end, we need the solution $\chi_0(t)$ of 
\begin{equation}
  A_{\mathrm{ref}} \chi_0=-\ddot{\chi}_0=0\;, \qquad \chi_0(0)=-b,\;\dot{\chi}_0(0)=a
\end{equation} 
which is given by
\begin{equation}
 \chi_0(t)=at-b\; .
\end{equation} 
Then 
\begin{equation}\label{eq:F0}
 F_{\mathrm{ref}}(0)=c\chi_0(T)+d\dot{\chi}_0(T)=acT-cb+da
\end{equation} 
which, of course, also follows from (\ref{eq:ha1}) for $\lam\to 0$. Combining (\ref{eq:detA0}) and (\ref{eq:F0}) we arrive at (\ref{eq:resdetA0}).


\section{Calculation of $\det A'$}\label{sec:appC}
In this appendix we sketch the calculation of the determinant of a Sturm-Liouville operator omitting its zero mode for the case of Robin boundary conditions, where we closely follow \cite{KiMcKa}. The main idea is to replace $F(\lam)$ as defined in (\ref{eq:defF}) by a function that is zero only at the {\em non-zero} eigenvalues of $A$ and behaves asymptotically in the same way as $F(\lam)$. Then all calculations may be done as before, and we end up with a relation similar to (\ref{eq:resdetA}). 

To get an idea how the replacement of $F(\lam)$ may look like, it is instructive to consider the scalar product between the zero-mode $\chi_0$ of $A$ and a general solution $\chi_\lam$ of (\ref{eq:odedetA}). Note that then $\chi_0$ fulfils {\em both} the boundary conditions at $t=0$ and $t=T$ whereas $\chi_\lam$ fulfils for general $\lam$ only the one at $t=0$ as specified in (\ref{eq:odedetA}). We find by partial integration 
\begin{align}
 \lam\la\chi_0|\chi_\lam\ra &= \la\chi_0|A\chi_\lam\ra \nn \\
  & = -\chi_0\dot{\chi}_\lam\rvert_0^T + \dot{\chi}_0\chi_\lam\rvert_0^T +\la A\chi_0|\chi_\lam\ra \nn \\
  & = -\frac{\chi_0(T)}{d}(c\chi_\lam(T)+d\dot{\chi}_\lam(T))
\end{align}
implying
\begin{equation}\label{eq:h1ab}
 F(\lam)=c\chi_\lam(T)+d\dot{\chi}_\lam(T)=-\frac{d}{\chi_0(T)}\lam\la\chi_0|\chi_\lam\ra\; .
\end{equation} 
Hence 
\begin{equation}
 \tilde{F}(\lam)=\frac{\lam-1}{\lam} F(\lam)=(1-\lam)\frac{d \;\la\chi_0|\chi_\lam\ra}{\chi_0(T)} \; ,
\end{equation} 
vanishes at all eigenvalues $\lam_n>0$ (since $F$ does), remains non-zero for $\lam=\lam_0=0$ (cf.~\ref{eq:h1ab}) and behaves asymptotically for large $\lam$ exactly as $F$. Similarly to the case without zero modes, one hence finds  
\begin{equation}
 \frac{\det A'}{\det A_{\mathrm{ref}}}=\frac{\tilde{F}(0)}{F_{\mathrm{ref}}(0)}
\end{equation}
which coincides with (\ref{eq:resdetAmod}).


\end{document}